\documentclass[a4paper]{jpconf}
\usepackage{amsmath}
\usepackage{graphicx}
\usepackage[numbers,sort&compress,square]{natbib}
\usepackage{url}

\usepackage{caption}
\usepackage{verbatim}
\usepackage{xcolor}
\newcommand{\rev}[1]{#1}% revision markup disabled for arXiv

\begin{document}

\title{Data-Driven Surrogate-Model for Real-Time Loads Prediction in Floating Wind Farms}

\author{Federico Bellini$^1$, Majid Ebrahimi$^1$, Alessandro Fontanella$^1$, 
Marcus Becker$^2$, Jan-Willem Van Wingerden$^2$, 
Sara Muggiasca$^1$, Marco Belloli$^1$}

\address{$^1$ Mechanical Engineering Department, Politecnico di Milano, 
Milano, Italy}
\address{$^2$ Delft Center for Systems and Control, TU Delft, Delft, 
Netherlands}

\ead{federico.bellini@polimi.it, alessandro.fontanella@polimi.it}

\begin{abstract}

This work presents a data-driven load-surrogate modelling framework for real-time control of floating offshore wind farms. The proposed methodology uses mid-fidelity simulations performed with OpenFAST and FAST.Farm together with MLife fatigue analysis to generate Damage Equivalent Load (DEL) datasets of floating wind turbines under wake steering conditions. A two-stage surrogate modelling approach is adopted: first, a Feed-Forward Neural Network (FFNN) is trained using data from simulations of a single floating wind turbine, while a second stage corrects the residual error between the FFNN predictions and FAST.Farm simulations using either Gaussian Process Regression (GPR) or an additional FFNN.
The results show that the proposed framework can estimate DELs with errors generally below 15\% for tower base bending moment, blade root bending moment, yaw bearing moment, and fairlead tension, while maintaining computational times compatible with real-time applications. The study also highlights the challenges related to generalization of the load-surrogate model under varying environmental and control conditions, especially for large floating wind farms. Overall, the proposed methodology demonstrates the potential of surrogate-based DEL estimation for load-aware floating wind farm control applications.

\end{abstract}

\vspace{1em}

\section{Introduction}
Wind energy is one of the key technologies driving the ongoing energy transition. Offshore wind, in particular, has the potential to accelerate this transition, as wind resources farther from the coast are generally stronger and more consistent. In these regions, where seabed depths typically exceed 60–70 m, bottom-fixed wind turbines become economically unattractive, motivating the development of floating offshore wind technology.

In recent years, both wind turbines and wind farms have grown significantly in size, with modern offshore wind farms comprising hundreds of turbines. As farm size increases, aerodynamic interactions between turbines become a major factor limiting overall power production. Upstream turbines generate wakes that reduce wind speed and increase turbulence intensity for downstream machines, leading to power losses and increased structural loading. To mitigate these wake-induced effects, several active control strategies have been proposed, including static approaches such as wake steering \cite{wake_steering_control, FLEMING2014211} and induction control \cite{induction_control, Bartl_2016}, and dynamic approaches such as dynamic induction control \cite{Pulse_control} and the helix strategy \cite{Helix_control}.

Significant effort has also been devoted to the development of low-fidelity wake modeling frameworks, such as FLORIS \cite{floris_model}, PyWake \cite{pywake2.5.0_2023}, and FLORIDyn \cite{Becker_2022}. These tools were originally designed to provide computationally efficient wind farm power predictions by simplifying the underlying flow physics, making them particularly suitable for AEP estimation and layout optimization including wake steering through yaw misalignment \cite{Bastankhah_Port-Agel_2016, wes-6-701-2021}. Subsequently, these frameworks have been extended to incorporate the effect of dynamic control approaches such as the helix strategy \cite{wes-2025-265, DAMMANN2026127808}. More recently, structural load considerations have been introduced to better capture the effect of these control strategies on turbine components, and load data resulting from wind farm simulations have been leveraged to develop surrogate models for Damage Equivalent Load (DEL) prediction under wake conditions \cite{Guillore_2024, wes-9-2039-2024, en17020346}.

The floating offshore wind sector is still at an early stage of deployment compared to bottom-fixed technology, with existing projects typically consisting of a limited number of turbines. Notable pilot projects include Hywind Scotland, a 30~MW farm comprising five spar-buoy turbines located off the coast of Peterhead, Scotland \cite{Equinor2023a}; WindFloat Atlantic, a 25~MW semi-submersible farm deployed 20~km off the coast of Viana do Castelo, Portugal \cite{PrinciplePower2024}; and Kincardine, a 50~MW farm consisting of five turbines installed 15~km off the coast of Aberdeen, Scotland \cite{Durakovic2021}. Additionally, the Floatgen demonstrator, a single barge-type floating wind turbine, has been operational off the coast of France since 2018 \cite{BWIdeol2024}. Some studies have investigated the relationship between wake interactions and structural loads on the main components of floating wind turbines. Carmo et al. \cite{wes-9-1827-2024} showed that the tower-base bending moment can increase by up to seven times compared to bottom-fixed turbines; however, no wind farm control strategy was considered in their study. Mahfouz \cite{Mahfouz_2023} developed a passive repositioning method based on an asymmetric mooring system to mitigate wake effects. With the same objective, Starink et al. \cite{Starink_2025} proposed a dynamic control strategy based on the repositioning of floating turbines through sinusoidal yaw motions. Moreover, Fontanella et al. \cite{wes-10-1369-2025} analyzed the effect of sinusoidal platform motions in the six degrees of freedom to accelerate wake recovery, although the implications for structural loads have so far received limited attention. More recently, Mohammadi et al. \cite{Mohammadi_we70111} investigated the structural effects of wake steering in a floating wind farm using OrcaFlex, showing that even under extreme operating conditions the structural integrity of the turbine was not compromised, further supporting the potential of floating wind technology.

Nevertheless, future commercial floating wind farms are expected to reach scales comparable to those of modern bottom-fixed installations, making wake mitigation strategies increasingly necessary. In this context, the structural impact of wind farm control strategies can no longer be neglected, and reliable tools for predicting structural loads within floating wind farms will become essential.

The models developed for bottom-fixed \cite{Guillore_2024, wes-9-2039-2024, en17020346} turbines cannot be directly applied to floating wind turbines because of their fundamentally different structural dynamics. The presence of floating platforms, mooring systems, and additional rigid-body motions introduces load transfer mechanisms and aero-hydro-servo-elastic couplings that are absent in bottom-fixed configurations, requiring dedicated modeling approaches. 
In this context, Singh et al. \cite{floating_load_surrogate_model} proposed a probabilistic surrogate model for DEL prediction on a spar-type floating wind turbine, demonstrating high accuracy across a range of metocean conditions, similarly, Yuan et al in \cite{jmse12081275} generated a dataset on a single floating turbine using openFAST and compared the performances of five machine learning algorithm for the training. However, their works focused on an isolated turbine and did not account for wake interactions within a wind farm environment. In floating wind farms, the coupling between aerodynamic wake effects and hydrodynamic loading introduces additional complexity, as platform motions induced by wind-wave interactions may amplify structural loads and influence the effectiveness of wake control strategies.

For these reasons, this paper proposes a methodology to estimate DELs on the most critical structural components of floating wind turbines accounting for aerodynamic interactions due to wakes while remaining computationally efficient enough for real-time applications. Such a methodology could be integrated into cost functions used for optimization of wake control strategies, enabling control strategies that account not only for power production but also for structural loading. Furthermore, due to the limited availability of large-scale floating wind farm simulations in the literature, this work also presents three large floating wind farm simulations used as a test set for the proposed load surrogate model. These simulations provide insight into the spatial and temporal variability of DELs over a 10-minute period across a large wind farm under synthetic wind conditions.

The main contributions of this work are as follows:
\begin{itemize}
    \item[(i)] Investigation of the possibility of integrating DEL computation within low-fidelity power calculation frameworks for floating wind farms;
    \item[(ii)] Development of a new methodology that reduces computational cost compared to currently available approaches;
    \item[(iii)] Comparison between Gaussian Process Regression (GPR) and Feed-Forward NNs (FFNN) for residual regression modeling;
    \item[(iv)] Large-scale floating wind farm simulations using FAST.Farm, including DEL computation for each wind turbine;
    \item[(v)] Assessment of the accuracy of the load surrogate model using MLife-based DEL evaluation.
    \item [(vi)] A demonstration of the surrogate model integrated within FLORIDyn, illustrating the trade-off between farm-level power production and structural DELs across a range of yaw misalignment angles for a representative environmental condition.
\end{itemize}

The remainder of the paper is organized as follows:
Section~\ref{sec:methodology} describes the proposed framework, including the modeling tools and dataset generation procedure.
Section~\ref{sec:results} presents the surrogate model predictions and compares them against large-scale wind farm simulations under three different operating conditions. 
Sections~\ref{sec:discussions} and~\ref{sec:conclusions} summarize the main findings and outline directions for future work.

%% =======================================================================
%% =======================================================================

%  Methodology

\section{Methodology}
\label{sec:methodology}

This section introduces the methodology developed to build the load surrogate model for floating wind turbines. Section~\ref{sec:reference_scenario} introduces the reference wind farm scenario used as the test case. Section~\ref{sec:wind_turbine_and_simulators} presents the tools and setup used to perform the FAST.Farm simulations composing the training dataset. Section~\ref{sec:SR_optimization} presents the yaw optimization strategy adopted for the large wind farm. Section~\ref{sec:model_architechture} presents the architecture used to build the DEL surrogate model. Finally, Section~\ref{sec:DEL_calculation} gives the main equation used for DEL calculation. 
%% =======================================================================
%% =======================================================================

\subsection{Reference scenario}
\label{sec:reference_scenario}
\begin{figure}[t]
\centering
\includegraphics[width=1\textwidth]{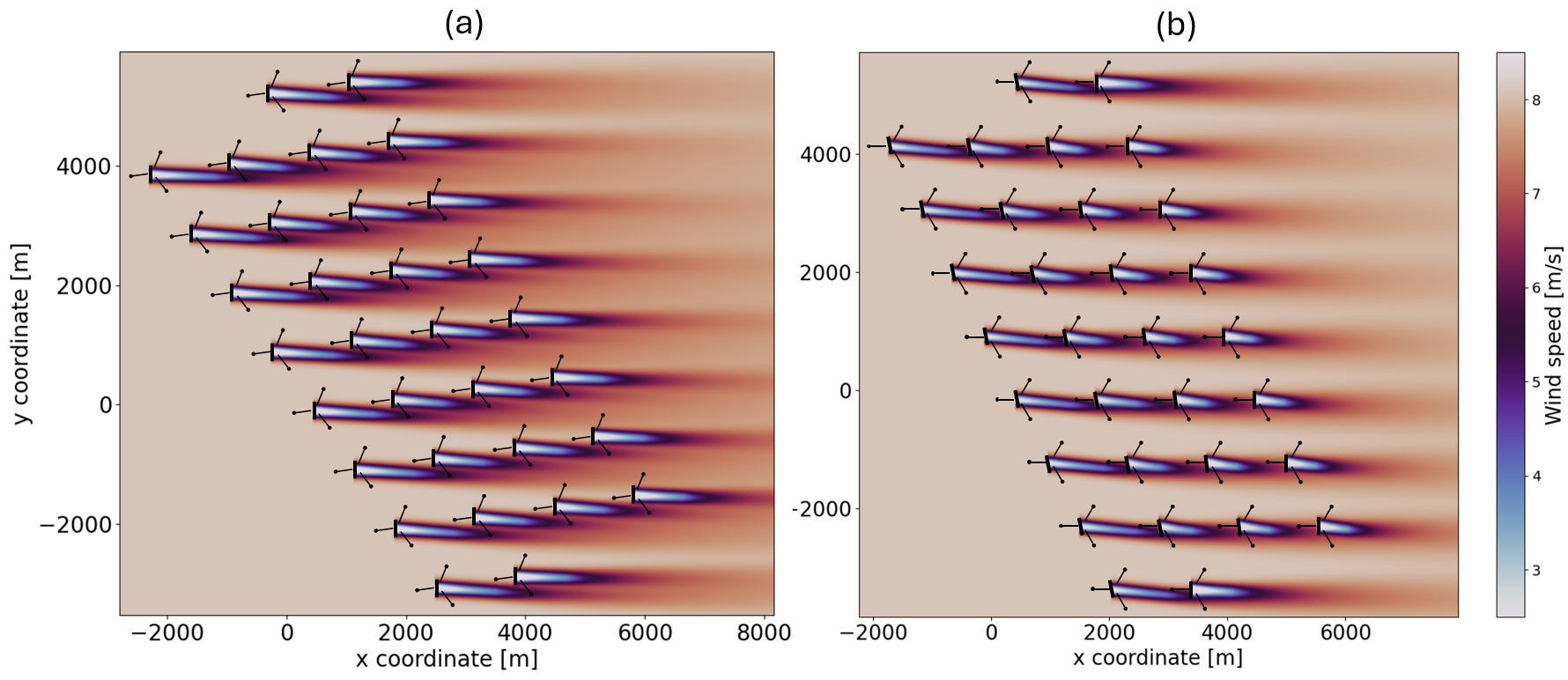}
\captionsetup{skip=2pt}
\caption{Wind farm layouts considered in this study: (a) staggered configuration and (b) fully aligned configuration.}
\label{fig:wind_farm_layouts}
\end{figure}

The reference scenario considers a floating wind farm of 32 turbines arranged according to the TotalControl reference wind farm layout \cite{TotalControl}, scaled to the rotor diameter of the IEA~15\,MW reference turbine \cite{gaertner2020iea,allen2020definition}. The layout features staggered rows with direction-dependent inter-turbine spacing: for north--south winds the streamwise spacing is approximately $5\,D$, whereas for east--west winds it increases to approximately $10\,D$. Figure~\ref{fig:wind_farm_layouts} illustrates the wind farm in the optimised-yaw configurations.
For wind from the north-west sector ($\approx297^\circ$ in the wind rose), the effective spacing projected along the wake propagation direction is approximately $5.6\,D$. The $5.6\,D$ configuration is the one adopted in the following analysis. Two cases are considered: the nominal $297^\circ$ direction and a case with an $8^\circ$ angular misalignment from it. Each wind turbine is an IEA~15\,MW offshore reference turbine mounted on the UMaine VolturnUS-S semi-submersible platform~\cite{gaertner2020iea,allen2020definition}.
Four large-scale wind farm simulations are performed to assess the generalisation capability of the surrogate model under environmental conditions not included in the training dataset. The four cases share the same ambient conditions: mean hub-height wind speed of 8\,m/s and turbulence intensity of 6\%. The wind speed is chosen in the below-rated region, where wake deficits are stronger and power losses are more significant since rotor thrust is higher and power production is not yet saturated. A turbulence intensity of 6\% is representative of typical offshore conditions at this wind speed, consistent with measurements reported in~\cite{TURK2010466,wes-9-2001-2024}. Irregular waves are modelled using a JONSWAP spectrum \cite{JONSWAP_spectra}. Two sea states are considered: a baseline condition matching one of the training dataset configurations ($H_\mathrm{s} = 2$\,m, $T_\mathrm{p} = 7$\,s, wave direction $= 0^\circ$), and an out-of-distribution condition ($H_\mathrm{s} = 3$\,m, $T_\mathrm{p} = 8$\,s, wave direction $= 5^\circ$) introduced to assess the generalization capability of surrogate model. The four simulation cases differ in wind direction, yaw control strategy, and sea state:

\begin{itemize}
    \item \textit{Aligned greedy}: wind direction aligned with the farm layout, 
          no wake steering, baseline sea state;
    \item \textit{Staggered greedy}: wind direction staggered by $8^\circ$ 
          with respect to the farm layout, no wake steering, baseline sea state;
    \item \textit{Staggered with wake steering}: same $8^\circ$ staggered wind 
          direction, yaw offsets optimised using FLORIS and the Serial Refine 
          algorithm~\cite{serial_refine}, baseline sea state;
    \item \textit{Staggered with wake steering, unseen sea state}: same as the previous case, but with the out-of-distribution sea state 
          ($H_\mathrm{s} = 3$\,m, $T_\mathrm{p} = 8$\,s, wave direction $= 5^\circ$).
\end{itemize}

%% =======================================================================
%% =======================================================================

\subsection{Wind turbine and wind farm simulators}
\label{sec:wind_turbine_and_simulators}
Wind farm simulations are performed using FAST.Farm~\cite{jonkman2021fast}, 
a mid-fidelity tool that couples the Dynamic Wake Meandering (DWM) \cite{Dynamic_wake_meandering} model for farm-level flow dynamics with OpenFAST \cite{openFAST} for single-turbine aero-servo-hydro-elastic computations. The wake is represented by a series of planes advected downstream from the rotor plane. Since FAST.Farm does not natively provide a model for vertical wake deflection, the curl wake model \cite{curl_model} is adopted in this work, as recommended by the developers. The 
low and high-resolution grid spacing and time step are chosen according 
to the recommendations in~\cite{jonkman2021fast} for the full range of 
simulated wind speed conditions.

Within OpenFAST, the following modules are activated. AeroDyn computes the aerodynamic loads on the rotor using blade element momentum (BEM) theory, with corrections for dynamic stall, skewed wake, and tip losses. ElastoDyn provides a multi-body structural dynamics model for the rotor, nacelle, and tower degrees of freedom. HydroDyn evaluates the hydrodynamic loading on the semi-submersible platform through potential flow theory combined with Morison drag terms, with irregular waves modelled by a JONSWAP spectrum \cite{JONSWAP_spectra}. MoorDyn is a lumped-mass mooring line model representing the three catenary lines that connect the platform to the seabed. Finally, ServoDyn acts as the turbine control interface and is coupled with the ROSCO controller~\cite{rosco} for blade pitch and generator torque regulation.

The ambient wind fields are generated using 
TurbSim~\cite{jonkman2006turbsim}. The software calculates time series of three-component wind velocity across a grid of discrete points with spatial discretisation consistent with the FAST.Farm low- and high-resolution grid requirements. To increase statistical robustness of the simulation outputs, each wind condition is repeated with five different random seeds. Each simulation covers 600\,s, preceded by a transient period sufficient to allow full wake development that is discarded from post process of simulation results. The transient time to discard is computed as a multiple of 600\,s and is given by Eq. \eqref{transient_time}:

\begin{equation}
\label{transient_time}
    T_{\text{time}} = \frac{\text{Distance}}{\text{Effective } V}
\end{equation}

where $T_\mathrm{time}$ is the transient time, and $\text{Effective } V$ is the effective wind speed recommended by the FAST.Farm guidelines for this operation. The distance required for the wake to develop is calculated as in Eq.\,\eqref{dist}:

\begin{equation}
\label{dist}
    \text{Distance} = n_T \cdot (X_D \cdot D) + 3D
\end{equation}

where $n_T$ is the number of turbines, $X_D$ is the downstream spacing considered in each simulation (see Tab.\,\ref{tab:dataset_stage2}, \ref{tab:features_farm}), and $D$ is the rotor diameter. The effective wind speed is computed following the FAST.Farm guidelines as:

\begin{equation}
\label{eff_ws}
    \text{Effective } V = V \cdot \left(1 - \frac{1}{6}\right)
\end{equation}

$T_\mathrm{time}$ is then rounded up to the nearest multiple of 600\,s.

%% =======================================================================
%% =======================================================================

\subsection{FLORIS - Yaw angles optimization.}
\label{sec:SR_optimization}
The low-fidelity wake model FLORIS (FLOw Redirection and Induction in Steady State)~\cite{floris_documentation, floris_git}, developed by NREL, is used in this work exclusively as a control-oriented tool to generate optimized yaw setpoints for the wake steering validation cases. FLORIS is a steady-state engineering framework for wind farm modeling and control, consisting of four main components: an empirical wake model describing velocity deficit, a deflection model capturing wake steering effects, a turbulence model accounting for wake-added turbulence, and a superposition model describing wake interactions in multi-turbine layouts. Among the available wake models, the Gauss–Curl Hybrid model is selected ~\cite{Bastankhah_Port-Agel_2016, Niayifar2016AnalyticalMO, wes-6-701-2021}. The model combines a Gaussian wake deficit formulation with an empirical representation of wake-induced vorticity and turbulence effects.

Yaw angle optimization is performed using the Serial Refine algorithm~\cite{serial_refine}, which provides a balance between computational efficiency and convergence robustness for steady-state wind farm control problems. The method relies on the assumption that the yaw setting of a downstream turbine does not affect the power of the turbines upstream of it, which allows the optimisation to be decoupled and solved sequentially. It proceeds in two stages. In the serial stage, the yaw range is discretised into a user-defined set of candidate angles; starting from the most upstream turbine, each candidate is applied in turn, the total farm power is evaluated, and the angle yielding the highest power is retained. This is repeated for each turbine moving downstream, with the most downstream turbine fixed at zero yaw, as it has no downstream wakes to steer. In the subsequent refine stage, the search interval is narrowed around the angle selected in the previous stage and the serial procedure is repeated, progressively increasing the angular resolution.
The number of refinement steps, the interval width, and its discretisation are user-defined parameters. In this work, three refinement steps were used: the initial yaw interval spanned $[-20^\circ, +20^\circ]$ discretised into 11 candidates, while each subsequent refinement halved the interval width and discretised it into 4 values.

%% =======================================================================
%% =======================================================================

\subsection {DEL surrogate model construction and generation of the training dataset}
\label{sec:model_architechture}
%In this subsection, the main modeling choices adopted in this work are justified. 
Floating wind turbine simulations require significantly higher computational cost compared to conventional bottom fixed wind turbine simulations because of the higher computational cost required to simulate the dynamic platform motion and the non linear mooring system. 
The generation of a load surrogate model from data requires a large number of wind turbine response data including component-level loads for the training process. 
Guillore et al.~\cite{Guillore_2024} built a load surrogate model from FAST.Farm simulations of bottom-fixed wind turbines, using rotor-sectorized turbulence intensity and wind speed as training features. In this work, a different methodology is adopted, inspired by Pish~\cite{Pish_2024}.
In \cite{Pish_2024}, the application differs from the present study, as the target quantity is the wake field rather than structural loads. However, the core idea is to decompose the modeling problem into two stages: a first stage in which the wake is reconstructed using high-fidelity single-turbine simulations, followed by a correction stage trained on a smaller number of full-farm simulations. In the present work, the same two-stage philosophy is applied to DEL prediction.

The present work assumes that turbulence intensity and mean wind speed capture a dominant contribution to fatigue load variability in wind turbine structures. This assumption is motivated by the well-established relationship between incoming turbulence and load fluctuations on key structural components, particularly in waked conditions \cite{DEL_turb}. While other effects such as wind shear, uncomplete wake overlap, and floating platform dynamics may also influence the structural response, they are not explicitly modeled in this framework and are partially accounted for through the correction stage of the surrogate model. 
The main idea to increase the amount of data available for the deep learning models is therefore to generate high-turbulence intensity wind conditions capable of mimicking waked wind turbine conditions, where turbulence intensity is naturally increased, and then correct the DEL prediction using a limited number of wind farm simulations while also considering the wind farm geometry and operational state. 

\begin{figure}[b]
\centering
\includegraphics[width=1\textwidth]{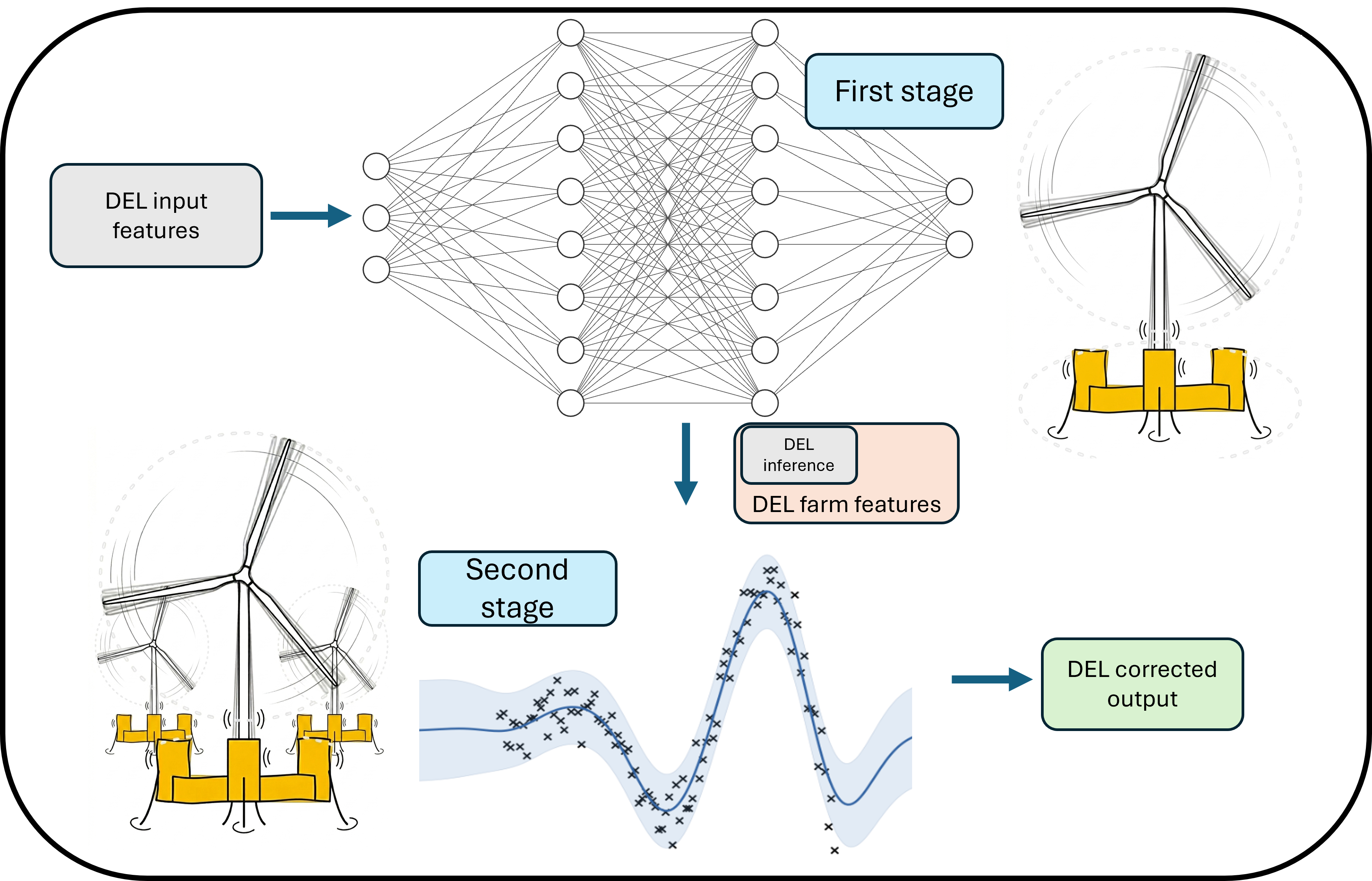}
\captionsetup{skip=2pt}
\caption{Framework utilized to build the DEL surrogate model.}
\label{fig:regression_architechture}
\end{figure}

This methodology, which splits the training process into two stages, enables the generation of a large amount of data using single-turbine simulations performed with OpenFAST, which are computationally much faster than simulating multiple floating wind turbines with FAST.Farm. The corresponding architecture is shown in Figure~\ref{fig:regression_architechture}.

\subsubsection{First stage load surrogate model}

The first stage consists of a regression model trained on a large coarse dataset of DEL values generated from isolated turbine simulations. The dataset generated for the single wind turbine simulations was created using the permutations of wind speed, wind direction, turbulence intensity, yaw angle, wave direction, and wind seeds, reported in Table~\ref{tab:dataset_stage1} for a total of 25515 simulations.

\begin{table}[h]
\centering
\caption{First stage training features used to generate the dataset.}
\label{tab:dataset_stage1}
\begin{tabular}{l | l}
\hline
\textbf{Parameter} & \textbf{Values} \\
\hline
$V$ [m/s]        & $\{4, 5, ..., 10\}$ \\
$V_\mathrm{d}$ [deg]        & $\{-10, 0, 10\}$ \\
$I_\mathrm{u}$ [\%]         & $\{4, 6, ..., 20\}$ \\
$Yaw$ [deg]         & $\{-20, -15, ..., 20\}$ \\
$W_\mathrm{d}$ [deg]   & $\{-20, 0, 20\}$ \\
$H_\mathrm{s}$ [m]         & $\{1.5, 3, 4.5\}$ \\
$T_\mathrm{p}$ [s]         & $\{6, 9, 12\}$ \\
$Seeds$ [-]         & $5$ \\
\hline
\end{tabular}
\end{table}

Among these simulations, the sea states were distributed in a deterministic manner in order to cover the operating conditions as uniformly as possible.
In particular, the wave conditions are drawn from the $3$-by-$3$ grid of significant wave height and period combinations $(H_s \in  \{1.5, 3.0, 4.5\}$ m,  $T_p \in \{6,9, 12\})$ s yielding $9$ distinct sea states indexed  $k \in \{1, ..., 9\}$.
For each fixed combination of wind speed, turbulence intensity, and wave direction, the $9$ nacelle yaw are mapped bijectively onto the 9 sea states as:

\begin{equation}
\label{eq_permutation1}
    k = 1 + (5(y-1) +\delta)   \bmod 9   
\end{equation}

where $y \in \{1, ..., 9\}$ is the yaw index and $\delta$ is an offset defined as:

\begin{equation}
\label{eq_permutation2}
    \delta = \left(13(i-1)+7(j-1)+3(d-1)\right) \bmod 9
\end{equation}

with $i, j, d$ respectively denoting the indices of wind speed, turbulence intensity, and wave direction. The multiplier 5 is coprime with 9, guaranteeing that the mapping constitutes a complete permutation. Each sea state appears exactly once across the $9$ yaw angles for every fixed $V$, $I_u$, $V_d$, Yaw and Wave Direction combination. The offset $\delta$ shifts this permutation across different wind conditions, ensuring that no systematic association between a specific angle of yaw and a specific sea state persists globally across the dataset. The coefficients $13$, $7$, and $3$ are arbitrary integers to ensure that $\delta$  takes sufficiently distinct values across the wind condition index combinations, avoiding degenerate periodicities in the permutation shift. As a result, each of the 9 sea states appears in exactly $189$ of the $1701$ wind states per wind direction, uniformly distributed across all yaw angles and wind conditions. This is then repeated over the three wind directions and the 5 wind seeds, for the total amount of 25515 simulations. This design prevents collinearity between the wave and wind parameters in the training dataset, allowing the surrogate model to learn the independent contribution of each variable to the predicted fatigue loads. The selected sea states and wind conditions are reported in Table \ref{tab:dataset_stage1}.

\subsubsection{Second stage load surrogate model}

To account for the effect of partial wake overlap in the wind farm, a second regression stage is introduced to learn the residuals between the first-stage predictions and the FAST.Farm wind farm simulations. The overall surrogate model can therefore be written as:

\begin{equation}
\label{DEL_sm}
    \hat{\text{DEL}}_{\text{SM}} = \hat{\text{DEL}}_{\text{FFNN}}(\mathbf{f}) + \varepsilon(\mathbf{x})
\end{equation}

where $\text{SM}$ denotes the surrogate model, $\mathbf{f}$ represents the features used in the first-stage regression (reported in Table~\ref{tab:dataset_stage1}), and $\mathbf{x}$ denotes the features employed in the second-stage residual model (reported in Table~\ref{tab:features_farm}). $\varepsilon(\mathbf{x})$ represents the corrective residual term. In the case where GPR is adopted, the residual model is expressed as:

\begin{equation}
\label{residual_GPR}
    \varepsilon(\mathbf{x}) \sim \mathcal{N}\!\left(\mu_{\text{GP}}(\mathbf{x}),\, \sigma^2_{\text{GP}}(\mathbf{x})\right), \quad \text{(FFNN+GPR)}
\end{equation}

Alternatively, when a neural network is used in the second stage, the residual model becomes:

\begin{equation}
\label{residual_FFNN}
    \varepsilon(\mathbf{x}) \approx \hat{\text{FFNN}}_{\text{res}}(\mathbf{x}) \quad \text{(FFNN+FFNN)}
\end{equation}

Compared to the first stage, the wind farm dataset was reduced to a total of 2700 simulations, due to the higher computational cost and the assumption that most of the physics is already learned during the first stage. Since each simulation includes four wind turbines, this provides a total of 10800 DEL values for the residual correction stage.

\begin{table}[h]
\centering
\caption{Second stage dataset matrix.}
\label{tab:dataset_stage2}
\begin{tabular}{l | l}
\hline
\textbf{Parameter} & \textbf{Values} \\
\hline
$V$ [m/s]  & $[5, 11, 2]$ \\
$V_\mathrm{d}$ [deg]        & $\{-10, 0, 10\}$ \\
$I_\mathrm{u}$ [\%]           & $\{4, 6, 8\}$ \\
$Yaw$ [deg]         & $\{-20, -15, ..., 20\}$ \\
$W_\mathrm{d}$ [deg]   & $\{-20, 0, 20\}$ \\
$X_D$ [-]         & $\{4, 6, 8\}$ \\
$Y_D$ [-]         & $\{-1, -0.5, ..., 1\}$ \\
$H_s$ [m]         & $\{2, 4\}$ \\
$T_p$ [s]         & $\{7, 10\}$ \\
$Seeds$ [-]         & $5$ \\
\hline
\end{tabular}
\end{table}

\begin{table}[b]
\centering
\caption{Input features used in FFNN and GPR surrogate models.}
\label{tab:features_farm}
\begin{tabular}{l | l}
\hline
\textbf{Feature} & \textbf{Description} \\
\hline
\texttt{DEL\_FFNN}          & FFNN single-turbine DEL prediction \\
\texttt{XD}               & Downstream distance \\
\texttt{YD}               & Lateral offset \\
\texttt{gamma\_upstream}  & Upstream turbine yaw misalignment [deg] \\
\texttt{yaw0\_deg}        & Local turbine yaw angle [deg] \\
$H_s$                     & Significant wave height [m] \\
$T_p$                     & Peak wave period [s] \\
$V_\mathrm{d}$            & Wave direction [deg] \\
\texttt{mean\_wind}       & Mean wind speed [m/s] \\
\texttt{TI\_wind}         & Turbulence intensity [\%] \\
\texttt{turb}             & Turbine index [-] \\
$\sin(V_\mathrm{d})$                & Sine of wind direction \\
$\cos(V_\mathrm{d})$                & Cosine of wind direction \\
\hline
\end{tabular}
\end{table}

Unlike the first training stage, where sea states are assigned deterministically for each wind condition and repeated identically across all five seeds, in the second stage the sea states are distributed across the five seeds as well. The second-stage dataset comprises 2\,700 simulations, corresponding to 540 unique wind condition combinations (excluding seeds) and 12 possible sea states combinations. Distributing sea states only across the 540 wind conditions --- as in the first stage --- would result in each sea state appearing only 45 times in the full dataset. By distributing them across seeds as well, each sea state is seen 225 times, providing a more uniform and statistically robust coverage of the sea state space given the smaller total number of simulations available.
This choice is motivated by the smaller total number of simulations available, which requires a more efficient coverage of the sea state space. 
The sea state grid consists of 12 combinations of $(H_\mathrm{s},\, T_\mathrm{p},\, \psi_\mathrm{wave})$, with $H_\mathrm{s} \in \{2,\, 4\}$ m, $T_\mathrm{p} \in \{7,\, 10\}$ s, and $\psi_\mathrm{wave} \in \{-20^\circ,\, 0^\circ,\, 20^\circ\}$, intentionally different from the first-stage values to introduce additional variety in the training data. For each block of fixed $(X_D,\, Y_D,\, V,\, I_u)$, a deterministic starting index is computed, and the five seeds receive five distinct sea states by stepping through the 12-combination grid with step size 5, which is coprime with 12, guaranteeing that all five sea states within a block are always distinct. The deterministic starting index is computed as:
\begin{equation}
\label{farm_distribution_1}
    s = 1 + \left(5\,b + c\right) \bmod 12
\end{equation}

where $b$ is a unique integer enumerating the $(X_D, Y_D, V, I_u)$ blocks and $c$ is an offset that varies with wind direction. The five seeds $k \in \{1,\dots,5\}$ are then assigned sea states as:
\begin{equation}
\label{farm_distribution_2}
    \ell_k = 1 + \left(s - 1 + 5(k-1)\right) \bmod 12
\end{equation}

For each combination of wind speed, turbulence intensity, downstream spacing, lateral offset, and wind seed, encoded by a unique index $h$, the yaw angle of the upstream turbine $T_1$ is drawn from the ordered set $\Gamma = \{-20^\circ, -15^\circ, \ldots, 0^\circ, \ldots, 15^\circ, 20^\circ\}$ via a linear congruential permutation as in \eqref{gam_seleciton}:

\begin{equation}
\label{gam_seleciton}
\gamma_{T_1}(h) = \Gamma_{f}, \qquad f = \left(a_\gamma\, i + b_\gamma\right) \bmod 9,
\end{equation}

with multiplier $a_\gamma = 5$, which is coprime with 9, guaranteeing a complete permutation of all nine values. The offset $b_\gamma$ differs across the three wind directions ($0^\circ$, $+10^\circ$, $-10^\circ$), ensuring that the yaw sequences are decorrelated across the three datasets.

After the upstream yaw selection, the downstream turbine yaw angles are assigned according to one of three patterns, selected deterministically:
\begin{itemize}
    \item 15\% of cases: $(A,\, A,\, A,\, 0^\circ)$ - all active turbines share the same 
    yaw angle;
    \item 15\% of cases: $(A,\, A,\, A-5^\circ,\, 0^\circ)$ - the third turbine yaw is reduced 
    by $5^\circ$;
    \item 70\% of cases: $(A,\, A-5^\circ,\, A-10^\circ,\, 0^\circ)$ - progressive yaw reduction moving 
    downstream.
\end{itemize}

Here, $A = |\gamma_{T_1}(i)|$ denotes the magnitude of the upstream yaw angle selected in Eq.~\eqref{gam_seleciton}. For the second and third patterns, the magnitudes assigned to the downstream turbines are computed as $\max(A-5^\circ,0^\circ)$ and $\max(A-10^\circ,0^\circ)$, respectively: clamping at zero prevents the magnitude from becoming negative for small values of $A$ (e.g., $A=5^\circ$), which would otherwise cause the yaw angle to change sign. The fourth turbine, $T_4$, always has zero yaw, since it has no downstream turbine whose wake it could steer. The sign of the yaw angle is instead determined by the lateral offset $YD$: positive for $YD>0$, negative for $YD<0$, and equal to the sign of the upstream yaw angle for $YD=0$. This convention ensures that the applied yaw always deflects wakes away from downstream turbines, consistent with physically meaningful wake-steering strategies.

These patterns are designed to reflect physically realistic yaw configurations for wake-steering control: progressive yaw reduction from upstream to downstream turbines is consistent with optimal wake-steering strategies, where upstream turbines benefit more from large yaw offsets while downstream turbines require smaller corrections. The three patterns ensure dataset diversity across uniform, partially-graded, and fully-graded yaw distributions, improving the generalization capability of the surrogate model.

\subsubsection{Surrogate model architecture.}
The two stages of the surrogate model were implemented using two different approaches. A diagram on how the model is built is shown in Fig. \ref{fig:regression_architechture}. The first-stage surrogate model consists of a FFNN with 7 input features, as presented in Table \ref{tab:dataset_stage1}, where the simulation seed is not included among the input variables, and two output neurons predicting, respectively, the mean DEL $\hat{\mu}$ and the associated log-variance $\log\hat{\sigma}^2$. The network comprises two fully connected hidden layers of 128 neurons each with rectified linear unit (ReLU) activation, where the neurons of the last hidden layer are connected to both output heads as in Figure \ref{fig:regression_architechture}. 

The first stage is trained by minimizing the Gaussian negative log-likelihood (NLL):
\begin{equation}
    \mathcal{L} = \frac{1}{N}\sum_{i=1}^{N} \frac{1}{2}
    \left[\log\hat{\sigma}_i^2 + \frac{(y_i - \hat{\mu}_i)^2}{\hat{\sigma}_i^2}\right]
\end{equation}
using the Adam optimizer \cite{kingma2015adam} with learning rate $10^{-3}$ and weight decay $10^{-4}$. The main hyperparameters – the number of neurons per layer and the learning rate – were tuned empirically by varying the neuron count in powers of two and testing selected learning rate values, retaining the configuration yielding the lowest validation error. Training runs for a maximum of 500 epochs with early stopping (patience of 30 epochs, i.e.\ training is halted if the validation NLL does not improve for 30 consecutive epochs) on the validation NLL, with 75\% of the data assigned to the training set and the remaining 25\% to the validation set.

%For more detail about Neural Network background we remind the reader to the following manuscript \cite{goodfellow2016deep}

The second stage of the model was built using two different approaches described in Sect. \ref{sec:model_architechture}. The first one used GPR (see Eq. \eqref{residual_GPR}), with the aim of obtaining both the residual correction and an estimate of the associated uncertainty. The overall uncertainty of the two-stage surrogate is obtained by combining the aleatoric uncertainty of the FFNN and the epistemic uncertainty of the GPR in quadrature, as in Eq.~\eqref{eq:total_sigma}:

\begin{equation}
    \sigma_{\mathrm{tot}} = \sqrt{\sigma_{\mathrm{FFNN}}^2 + \sigma_{\mathrm{GPR}}^2}
    \label{eq:total_sigma}
\end{equation}

The second approach used an additional neural network model (see Eq. \ref{residual_FFNN}). The motivation for pursuing these two different strategies was twofold. First, the goal was to compare the accuracy of two widely used machine learning approaches. Second, both training and inference times are strongly affected by the number of input samples, particularly for GPR, whose training complexity scales as $\mathcal{O}(n^3)$, considerably worse than the near-constant inference time of the FFNN. During inference, this effect is less pronounced but still relevant. Neural networks, on the other hand, do not provide direct uncertainty estimation, but both training and inference times scale more favorably with increasing dataset size.

The GPR model employs an automatic relevance determination (ARD) kernel of the form:
\begin{equation}
\label{GPR_kernel}
    k(\mathbf{x}, \mathbf{x}') = \sigma^2 \exp\!\left(-\sum_{i=1}^{d} 
    \frac{(x_i - x'_i)^2}{2\ell_i^2}\right) + \sigma_n^2\,\delta(\mathbf{x},\mathbf{x}')
\end{equation}
where $\sigma^2$ is the signal variance, $\ell_i$ is the length scale associated with the $i$-th feature, and $\sigma_n^2$ is the noise variance estimated by a WhiteKernel, an additive kernel component that models independent Gaussian noise on the training observations. The ARD formulation assigns an independent length scale to each of the 13 input features, allowing the model to identify which features are most informative for the residual correction -- features with smaller $\ell_i$ contributing more strongly to the prediction. Input features are standardised prior to GPR fitting by subtracting the mean and dividing by the standard deviation of each feature computed on the training set (zero mean, unit variance). Kernel hyperparameters are optimised by maximising the log marginal likelihood with 5 random restarts. The model is trained on the residuals $\varepsilon = \hat{\text{DEL}}_{\text{FFNN}} - \text{DEL}_{\text{true}}$ using a row-shuffle split with a test size of 0.20.
%For more detail about GPR background we remind the reader to the following manuscript \cite{rasmussen2006gaussian}

The second-stage residual correction based on an additional neural network employs a simpler architecture with respect to the first-stage model, consisting of two fully connected hidden layers of 256 neurons each with ReLU activation and a single output neuron predicting the residual $\varepsilon$. The 
model takes as input the 13 features shown in Table \ref{tab:features_farm} and is trained by minimizing the Mean Squared Error (MSE) loss on standardized residuals using the Adam optimizer with learning rate $10^{-3}$, for a maximum of 300 epochs with early stopping (patience $= 30$) on the validation MSE, using a row-shuffle split with validation set using 20\% of dataset. Unlike the first-stage model, no uncertainty estimation head is included, as the FFNN+FFNN model serves primarily as a deterministic baseline for comparing residual correction accuracy against the FFNN+GPR framework, which natively provides predictive uncertainty through the posterior variance of the Gaussian process.

%% =======================================================================
%% =======================================================================

\subsection{Fatigue load calculation}
\label{sec:DEL_calculation}
This section introduces the methodology of short-term DEL analysis. DEL converts timeseries of fluctuating loads into constant-amplitude fatigue loads, using the rain-flow counting algorithm. The rain-flow cycle counting algorithm is widely used for fatigue analysis. 

MLife \cite{mlife_theory} is utilized to calculate fatigue damage, taking into account short-term DELs. MLife processes fluctuating loads by dividing them into individual cycles. Based on Miner’s rule, fatigue damage accumulates linearly, with each cycle characterized by a mean value and a range.

The Palmgren-Miner rule calculates the total damage at a given location subject to a fluctuating load as in Eq. \eqref{f1}:

\begin{equation}
\mathcal{D} = \sum_{i}^{} \frac{n_i} {N_i L_i^{RF}}
\label{f1}
\end{equation}

where $D$ is the cumulative fatigue damage, $n_{i}$ is the number of load cycles in block $i$, $N_{i}$ is the number of cycles before failure, and $L^{RF}_{i}$ is the cycle’s load range around a fixed mean value. According to the $S-N$ curve, the relationship between load range $L^{RF}_{i}$ and the cycles to failure $N_{i}$ is shown as follows in Eq. \eqref{f2}:
 
\begin{equation}
N_i = \Biggl( \frac{L^{ult}} {\frac{1}{2}L^{RF}_i} \Biggl)^m
\label{f2}
\end{equation}

where $L^\mathrm{ult}$ denotes the ultimate design load of the considered component and $m$ represents the W\"ohler exponent. W\"ohler exponent is 4 for structures made of steel (tower, mooring lines, yaw bearing) and 10 for structures made of composite (blade) \cite{Krathe2025}. The actual load cycles occur over a spectrum of load means, and thereby requires a correction to the fatigue cycles’ load ranges. The Goodman correction presented in Eq. \eqref{f3} is utilized to modify the load ranges of the fatigue cycles $L^{RF}_i$ to treat the data as if each cycle occurred around a fixed mean load.
 
\begin{equation}
L^\mathrm{RF}_i = L^R_i\Biggl( \frac{L^{ult}} {L^{ult}- L_i^{M}} \Biggl)
\label{f3}
\end{equation}

where $L^{R}_i$ denotes the load range of i-th cycles’s range about a load mean of $L^{M}_i$. For each time-series dataset, Eq. \eqref{f4} calculates the short-term fatigue damage equivalent at a frequency of 1 Hz, which results in the same level of damage as the corresponding stochastic load.

\begin{equation}
\mathrm{DEL}^\mathrm{ST} = \Biggl( \frac{\sum_{i}(n_{i}(L_{i}^{RF})^m)} {n^{STeq}} \Biggl)^{1/m}
\label{f4}
\end{equation}
 
\begin{equation}
n^{STeq} = f^{eq} T 
\label{f5}
\end{equation}
 
where $n_{i}$ is the number of cycles in the $i$-th bin of the rainflow histogram, $L^{RF}_{i}$ is the load range of cycles of a fixed load-mean value using the Goodman correction, $n^{STeq}$ is the total equivalent 
fatigue counts for the time series, $f^{eq} = 1$\,Hz.

To account for load directionality, the DELs are computed along several directions $\theta$ within the blade-root, tower-base, and nacelle bearing cross-section, $M(t, \theta)$, as in Eq. \eqref{f6}
 
\begin{equation}
M(t, \theta) = M_x(t)cos(\theta) + M_y(t)sin(\theta)
\label{f6}
\end{equation}
where $M_x$ and $M_y$ are the moments along two orthogonal directions. In this work we assume $\theta = 60^\circ$, and we consider the maximum DEL within these load roses.

\subsection{Demonstration of the load surrogate model implementation in FLORIDyn}
\label{sec:FLORIDyn_demonstration}

FLORIDyn ~\cite{Becker_2022} is a control-oriented dynamic wake model that adopts a Lagrangian 
description of the flow: at each time step, so-called Observation Points 
are generated at the rotor plane and propagate downstream carrying turbine 
states (e.g. yaw angle, axial induction), flow field states (e.g. wind 
speed, wind direction, turbulence intensity) and their own positional 
states.

In this work, the load surrogate model presented in Section \ref{sec:model_architechture} is integrated within FLORIDyn to enable online DEL prediction alongside the wake and power estimation. The purpose of this integration is to demonstrate the potential of the surrogate model as a real-time virtual sensor for structural loads within a control-oriented wind farm simulation framework, providing the load information required by load-aware wind farm control strategies at a computational cost compatible with online applications.

At each simulation step, FLORIDyn provides the effective wind speed and the added turbulence intensity at each turbine, which are combined with the imposed yaw setpoints and the prescribed sea-state conditions (fixed a priori, see Table \ref{tab:env_cond}) and fed as inputs to the surrogate model. The layout considered in the application consists of three floating wind turbines slightly misaligned with respect to the incoming wind direction of $265^\circ$.

\begin{table}[h]
\centering
\caption{Environmental conditions used in the FLORIDyn demonstration.}
\label{tab:env_cond}
\begin{tabular}{c c c c c c c}
\hline
$V$ [m/s] & $W_\mathrm{d}$ [deg] & $I_\mathrm{u}$ [\%] & $H_s$ [m] & $T_p$ [s] & $V_\mathrm{d}$ [deg] & $\gamma$ [deg] \\
\hline
$8$ & $265$ & $6$ & $3$ & $8$ & $5$ & $\{-20, -18, \dots, 0\}$ \\
\hline
\end{tabular}
\end{table}

%% =======================================================================
%% =======================================================================

%  Results

\section{Results}

\label{sec:results}
This section reports: (i) the accuracy of the single-turbine FFNN trained and validated on the full dataset of single-wake simulations (Sect.~\ref{subsec:First_stage_results}); (ii) the generalisation capability and prediction accuracy of both surrogate frameworks --- FFNN+GPR and FFNN+FFNN --- assessed on large-farm operating conditions unseen during training, including a case with a different sea state (Sect.~\ref{subsec:Load_surrogate_model_assessment}); (iii) the assessment of the uncertainty estimated by the FFNN+GPR load surrogate model (Sect.~\ref{subsec:uncertainty_assessment}); (iv) a demonstration of the surrogate model integrated within FLORIDyn 
(Sect.~\ref{subsec:demo}); (v) the computational inference time of both surrogate frameworks, comparing the FFNN and GPR second-stage 
implementations (Sect.~\ref{subsec:time_costs}).

\begin{figure}
\centering
\includegraphics[width=1\textwidth]{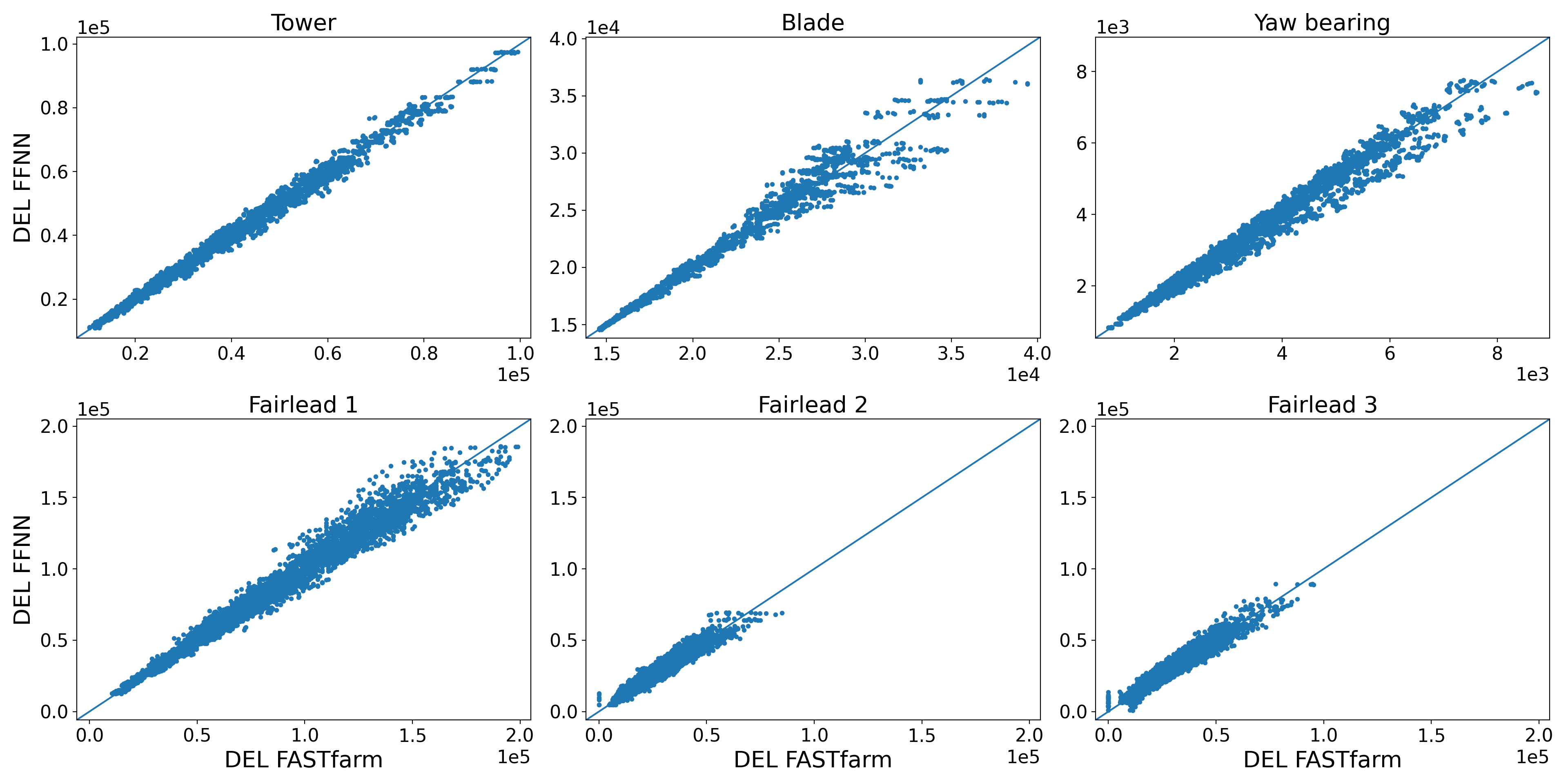}
\captionsetup{skip=2pt}
\caption{Identity plots coming from validation set of the first regression stage performed with FFNN of the training set for all the analyzed components.}
\label{fig:parity_vals}
\end{figure}

\subsection{Single-turbine FFNN evaluation}
\label{subsec:First_stage_results}
As a first step in the development of the load surrogate model, the accuracy of the first-stage FFNN was assessed. Overall, the FFNN achieves low prediction errors on the dataset and provides reliable uncertainty confidence intervals for all considered load components. The parity plots shown in Figure \ref{fig:parity_vals} indicate good agreement between predicted and reference values on the validation set across all load channels.

The only noticeable dispersion is observed for the blade-root DEL at higher load values. A possible explanation is that high DELs values – where the FFNN accuracy is lower – correspond to wind speeds near rated where the thrust force along the blades is maximum. Under these near-rated operating conditions, blade pitch activity becomes more pronounced as the controller regulates the turbine power output. Consequently, the underlying physical behaviour of the wind turbine becomes more complex due to the stronger coupling between aerodynamic loading and blade-pitch control actions. This increased complexity may not be fully captured by the FFNN, leading to the larger prediction scatter observed at high DEL values. Furthermore, some of the reference DEL values calculated in FAST.Farm are equal to zero. These correspond to cases in which MLife computes the DEL under very low load-signal variability. In such conditions, MLife applies an internal filtering criterion that results in a zero DEL value. Interestingly, not all DELs obtained from the same simulation under different turbulence seeds are zero, although the remaining values are still considerably low. Additional considerations regarding the lower importance of Fairlead tensions 2 and 3 will be discussed in the following sections.

\subsection{Load surrogate model assessment}
\label{subsec:Load_surrogate_model_assessment}
The second-stage residual models --- both FFNN+FFNN and FFNN+GPR --- achieve similar levels of accuracy on the validation set, as summarised in Table~\ref{tab:accuracy}. The main practical difference between the two approaches lies in computational cost: the FFNN residual model trains in a matter of minutes for all six loads components, whereas the GPR requires approximately one day of training time due to its $\mathcal{O}(n^3)$ complexity, where $n$ is the number of training samples.
Although slower to train, the GPR natively provides an uncertainty estimate through its posterior variance, which is not available in the FFNN+FFNN framework. The GPR length scales reported in Table~\ref{tab:kernel} provide insight into the relative importance of each feature for the residual correction. A small length scale indicates that the residual varies rapidly with respect to that feature, meaning the GPR relies heavily on it during inference. Conversely, a large length scale indicates that the feature has little influence on the predicted residual. As an example, for the blade bending moment the residual 
correction relies primarily on two features: the first-stage DEL prediction ($\ell = 0.060$) and the mean wind speed 
($\ell = 0.075$). All other features exhibit length scales at least one order of magnitude larger, indicating that 
variations in those features have negligible influence on the residual prediction. The implications of this behaviour 
for large-farm generalisation are discussed later in this section,.

\begin{table}[h]
\centering
\caption{Surrogate model accuracy (PctMAE).}
\label{tab:accuracy}
\begin{tabular}{l | c c c}
\hline
\textbf{Component} & \textbf{FFNN [\%]} & \textbf{FFNN+FFNN [\%]} & \textbf{FFNN+GPR [\%]} \\
\hline
Tower        & 11.43 & 4.34 & 3.57 \\
Blade        & 5.51  & 1.95 & 1.97 \\
Bearing      & 7.42  & 4.08 & 3.68 \\
Fairlead 1   & 9.42  & 6.30 & 5.78 \\
Fairlead 2   & 15.00 & 9.55 & 8.96 \\
Fairlead 3   & 16.04 & 9.82 & 9.29 \\
\hline
\end{tabular}
\end{table}

\begin{figure}[t]
\centering
\includegraphics[width=1\textwidth]{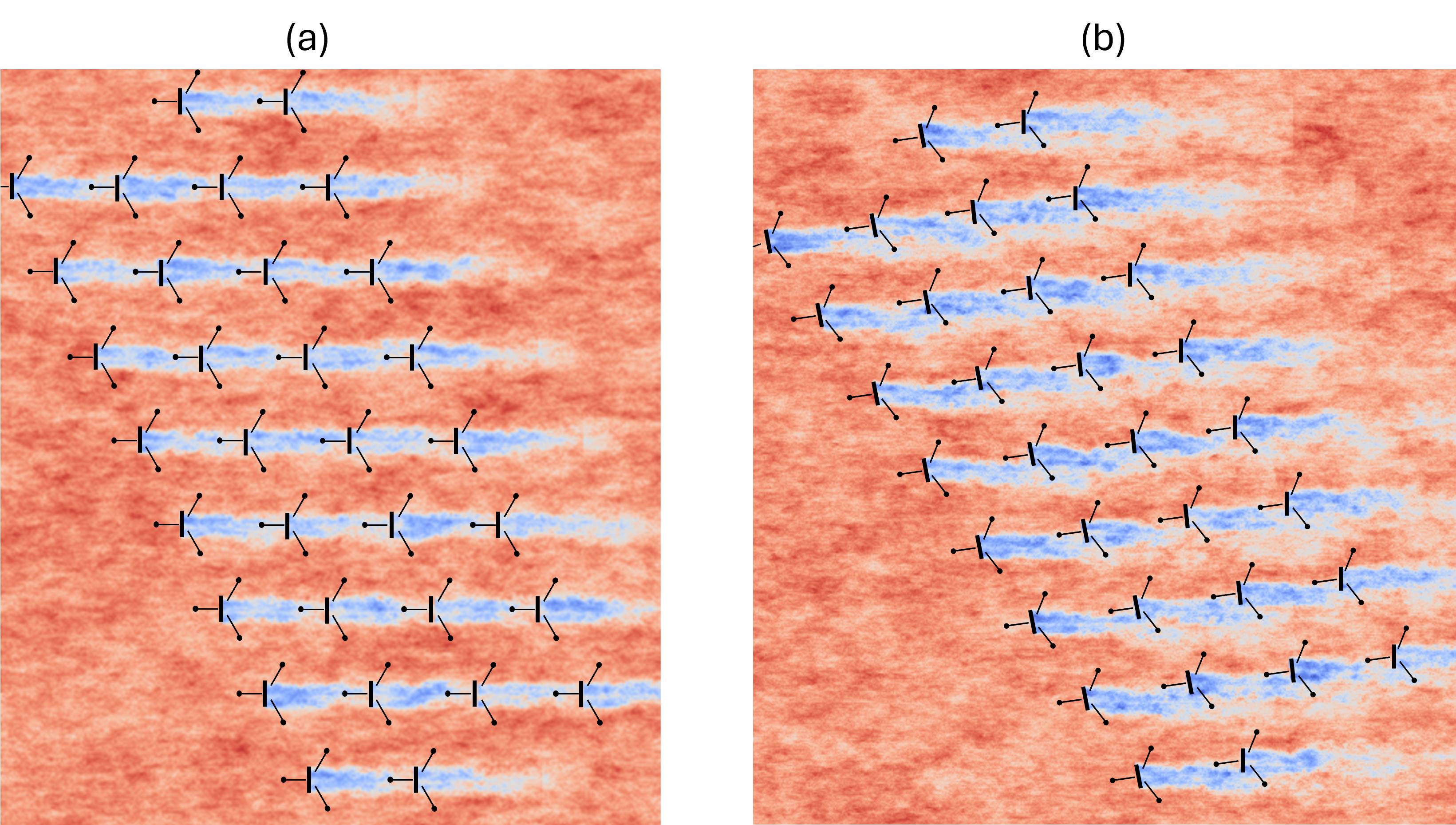}
\captionsetup{skip=2pt}
\caption{Instantaneous wind field across the wind farm from FAST.Farm for a mean wind speed of 8 m/s and a turbulence intensity of 6\%. (a) The mean wind direction is aligned with the turbine rows. (b) The mean wind direction forms an angle of $8^\circ$ with respect to the turbine rows.}
\label{fig:Large_farm}
\end{figure}

\begin{figure}[t]
\centering
\includegraphics[width=1\textwidth]{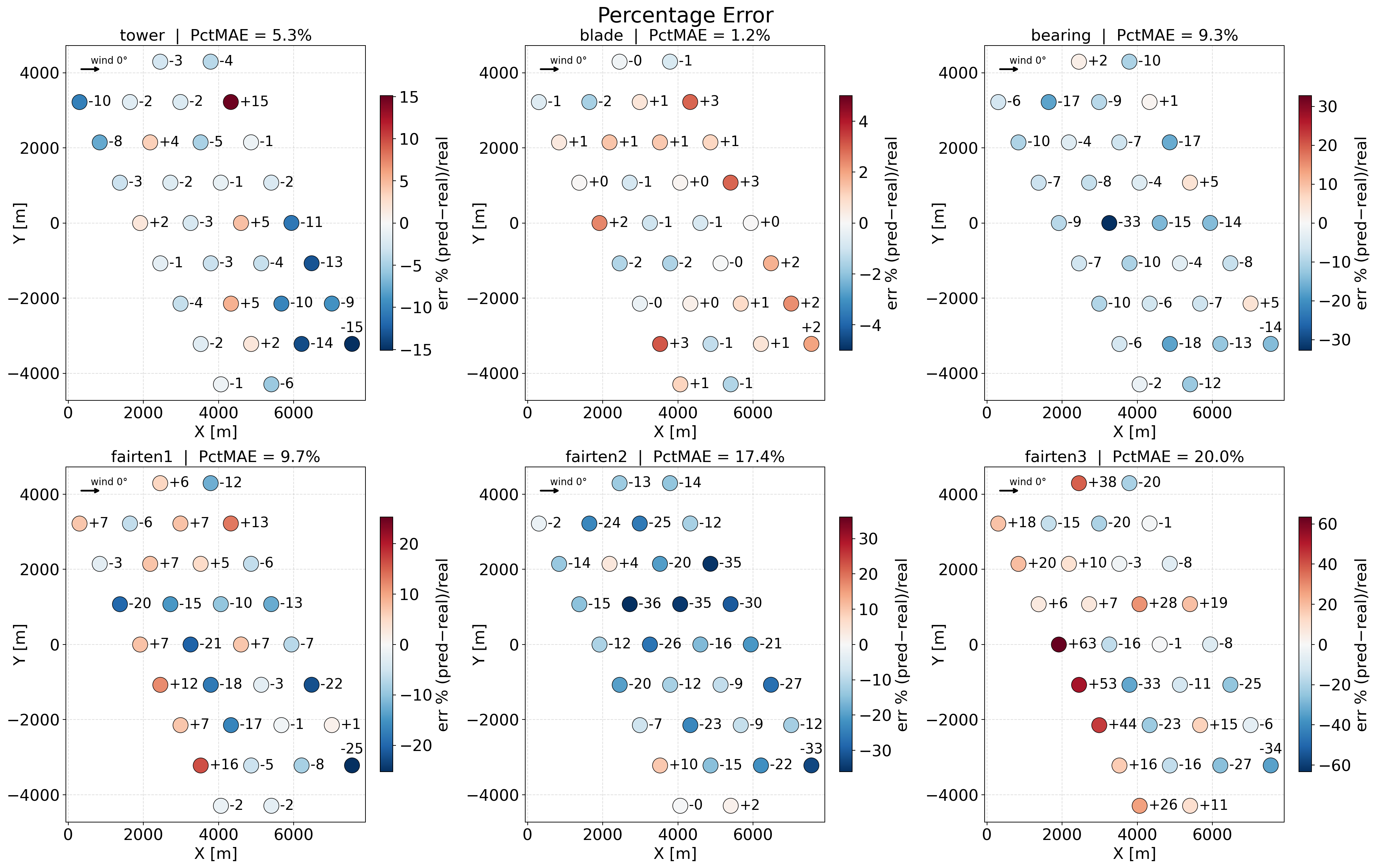}
\captionsetup{skip=2pt}
\caption{Full aligned layout wind farm showing the percentage error between the real DEL and the surrogate model.}
\label{fig:error_26d6}
\end{figure}

As expected, the first-stage FFNN output (DEL\_FFNN) is among the most influential features for most components, confirming that the first stage of the surrogate model captures the dominant physical behaviour of the loads responses and that the residual correction operates primarily on top of it. 
For the blade load component, DEL\_FFNN has a length scale of 0.060 --- by far the smallest across all features --- indicating that the GPR correction is almost entirely driven by the first-stage prediction. In this situation, the GPR effectively interpolates the residual as a function of DEL\_FFNN rather than learning a smooth correction across the full feature space. As a consequence, when the model is applied to the large farm simulations --- where the first stage produces DEL\_FFNN values that fall outside the training distribution --- the GPR finds no reliable support points and its correction degrades. This explains why, for the blade load, the residual correction worsens the overall prediction in the large farm cases, and why retaining only the first-stage model yields more robust results, with errors below 5\% which is considered satisfactory for real-time DEL estimation (Table \ref{tab:comparison_all}). The wind farm layout analyzed is presented in Figure \ref{fig:Large_farm} and the single errors in the predictions averaged in Table \ref{tab:comparison_all} are also shown turbine by turbine in Figure \ref{fig:error_26d6}

The opposite behaviour is observed for the tower bending moment, where all features exhibit length scales of comparable magnitude. This indicates that the GPR learns a smooth correction distributed across the full feature space, enabling better generalisation to unseen conditions. Accordingly, the tower bending moment DEL predictions consistently improve across all three large-farm simulations, reducing the error in a range between 5\% and 10\%.

For the bearing and fairlead~1, an intermediate behaviour is observed:  a few features dominate the correction while others have negligible influence, resulting in modest improvements that are not consistent across all tested operating conditions --- in some cases the residual correction does not reduce the prediction error compared to the first-stage model alone. It is worth noting that features with very large length scales --- such as YD for the tower or TI for the blade --- contribute negligibly to the residual prediction and could in principle be removed from the second-stage model without affecting inference accuracy, reducing training cost. 

The first two columns of Table~\ref{tab:kernel} report the individual kernel parameters $\sigma^2$ and $\sigma_n^2$. Another useful metric to assess the goodness of the training is their ratio, $\sigma^2/\sigma_n^2$, which quantifies how much of the residual variance is explained by structured patterns in the feature space ($\sigma^2$) relative to unexplained noise ($\sigma_\mathrm{n}^2$). For the tower, $\sigma^2/\sigma_n^2 \approx 6.9$, indicating that the GPR operates in a signal-dominated regime. For fairlead~1, 2, and 3, this ratio drops to approximately 3.5, 2.8, and 4.5, respectively, suggesting that a significant fraction of the residual variability cannot be explained by the available features. This is consistent with the hypothesis that mooring-line loads are governed by platform motions, which are not explicitly included in the feature set of the current surrogate model.

Analysing the behaviour of the FFNN+FFNN residual model is inherently more difficult than for the GPR case, as FFNNs offer limited interpretability compared to GPR. Nevertheless, the results reported in Table~\ref{tab:comparison_all} show that the FFNN+FFNN framework achieves accuracy comparable to FFNN+GPR across most load components and tested operating conditions, suggesting that the residual model based on FFNN captures correction patterns similar to the GPR without an explicit representation of the feature space. 
A notable exception is the blade bending moment, where FFNN+FFNN appears to suffer less from out-of-distribution degradation than FFNN+GPR on the large farm simulations, which is consistent with the GPR limitations discussed above.

\begin{table}[h]
\centering
\caption{Fitted GPR kernel parameters. RBF length scales per feature.}
\label{tab:kernel}
\resizebox{\textwidth}{!}{%
\begin{tabular}{l | c c | c c c c c c c c c c c c c}
\hline
\textbf{Comp.} & $\boldsymbol{\sigma}$ & $\boldsymbol{\sigma_n}$ &
\texttt{DEL\_FFNN} & \texttt{DD} & \texttt{YD} & $\boldsymbol{\gamma}$\texttt{\_up} & \texttt{yaw0} &
$\boldsymbol{H_s}$ & $\boldsymbol{T_p}$ & \texttt{wvdir} & \texttt{WS} & \texttt{TI} &
\texttt{turb} & $\boldsymbol{\sin\psi}$ & $\boldsymbol{\cos\psi}$ \\
\hline
Tower      & 0.711 & 0.271 & 0.586 & 1.27  & 100   & 16.3 & 15.9 & 0.422 & 0.435 & 2.59  & 0.425 & 0.597 & 0.882 & 2.11  & 3.20  \\
Blade      & 1.28  & 0.061 & 0.060 & 4.49  & 0.386 & 32.0 & 2.59 & 5.72  & 20.1  & 7.13  & 0.075 & 100   & 1.65  & 16.7  & 4.06  \\
Bearing    & 1.04  & 0.470 & 0.572 & 2.10  & 10.9  & 25.9 & 7.63 & 0.986 & 1.68  & 1.04  & 0.502 & 1.29  & 1.59  & 3.56  & 2.36  \\
Fairlead 1 & 0.816 & 0.435 & 0.504 & 1.38  & 7.96  & 5.49 & 7.95 & 0.966 & 0.925 & 1.82  & 0.304 & 0.549 & 0.880 & 1.77  & 3.25  \\
Fairlead 2 & 0.872 & 0.519 & 0.872 & 6.38  & 2.35  & 8.14 & 4.56 & 1.03  & 1.62  & 0.881 & 0.398 & 1.16  & 3.36  & 2.07  & 1.93  \\
Fairlead 3 & 1.24  & 0.588 & 1.22  & 5.50  & 2.04  & 17.4 & 3.84 & 1.21  & 2.76  & 1.32  & 0.980 & 3.11  & 3.82  & 5.13  & 1.34  \\
\hline
\end{tabular}}
\end{table}

%% ========================== OPZIONE TABELLE ACCOPPIATE =============================

\begin{table}[h]
\centering
\caption{Percentage MAE of the standalone FFNN and residual-corrected FFNN+GPR and FFNN+FFNN frameworks under four operating conditions: aligned (greedy), misaligned greedy, misaligned with optimized yaw, and misaligned with optimized yaw and different sea state.}
\label{tab:comparison_all}
\resizebox{\textwidth}{!}{%
\begin{tabular}{l | ccc | ccc | ccc | ccc}
\hline
& \multicolumn{3}{c|}{\textbf{Aligned (greedy)}}
& \multicolumn{3}{c|}{\textbf{Misaligned (greedy)}}
& \multicolumn{3}{c|}{\textbf{Misaligned (yaw opt.)}}
& \multicolumn{3}{c}{\textbf{Misaligned (yaw opt., ss)}} \\
\cline{2-13}
\textbf{Target} & FFNN & +GPR & +FFNN & FFNN & +GPR & +FFNN & FFNN & +GPR & +FFNN & FFNN & +GPR & +FFNN \\
\hline
Tower      & 16.03 & 5.52  & 5.31  & 11.998 & 5.744  & 6.245  & 13.151 & 6.592  & 6.568  & 16.01 & 8.20  & 9.92  \\
Blade      & 1.56  & 6.32  & 1.19  & 4.558  & 13.554 & 7.536  & 3.067  & 10.235 & 5.447  & 2.61  & 9.47  & 5.31  \\
Bearing    & 9.85  & 10.58 & 9.33  & 8.761  & 7.543  & 10.512 & 9.283  & 9.084  & 9.615  & 10.04 & 8.09  & 5.58  \\
Fairlead 1 & 12.65 & 13.04 & 9.74  & 8.265  & 8.854  & 11.333 & 8.005  & 8.492  & 8.150  & 8.59  & 6.73  & 9.17  \\
Fairlead 2 & 16.33 & 20.84 & 17.38 & 23.748 & 25.162 & 23.810 & 19.076 & 20.171 & 22.267 & 15.71 & 20.38 & 20.52 \\
Fairlead 3 & 14.60 & 17.94 & 20.04 & 15.308 & 18.949 & 22.458 & 36.662 & 36.380 & 37.881 & 28.77 & 23.91 & 28.26 \\
\hline
\end{tabular}%
}
\end{table}

Despite these limitations, the surrogate model is generally able to estimate DELs with an error below 15\% for the tower base bending moment, blade root bending moment, yaw bearing bending moment, and fairlead~1 tension across all four simulated operating conditions of the test set. The accuracy for fairlead~2 and fairlead~3 remains comparatively poor, with errors exceeding 20\% in several cases. However, the load amplitudes experienced by fairlead 2 and fairlead 3 are typically considerably lower than those of fairlead 1, as also shown in the bottom plots in Fig. \ref{fig:parity_vals}. According to Eq.~\eqref{f2}, the number of cycles to failure $N_i$ scales as $(L_i^{RF})^{-m}$; with the S--N exponent used in this study ($m = 4$), considering, as an example, a five-fold reduction in load range corresponds to an increase in $N_i$ by a factor of approximately $5^4 \approx 625$, i.e., more than two orders of magnitude. As a result, the contribution of fairlead 2 and fairlead 3 to the cumulative fatigue damage $D$ in Eq.~\eqref{f1} is substantially reduced relative to fairlead 1. Therefore, the lower prediction accuracy observed for these two mooring lines is expected to have a minor influence on the overall fatigue assessment of the platform.

\begin{figure}[ht]
\centering
\includegraphics[width=1\textwidth]{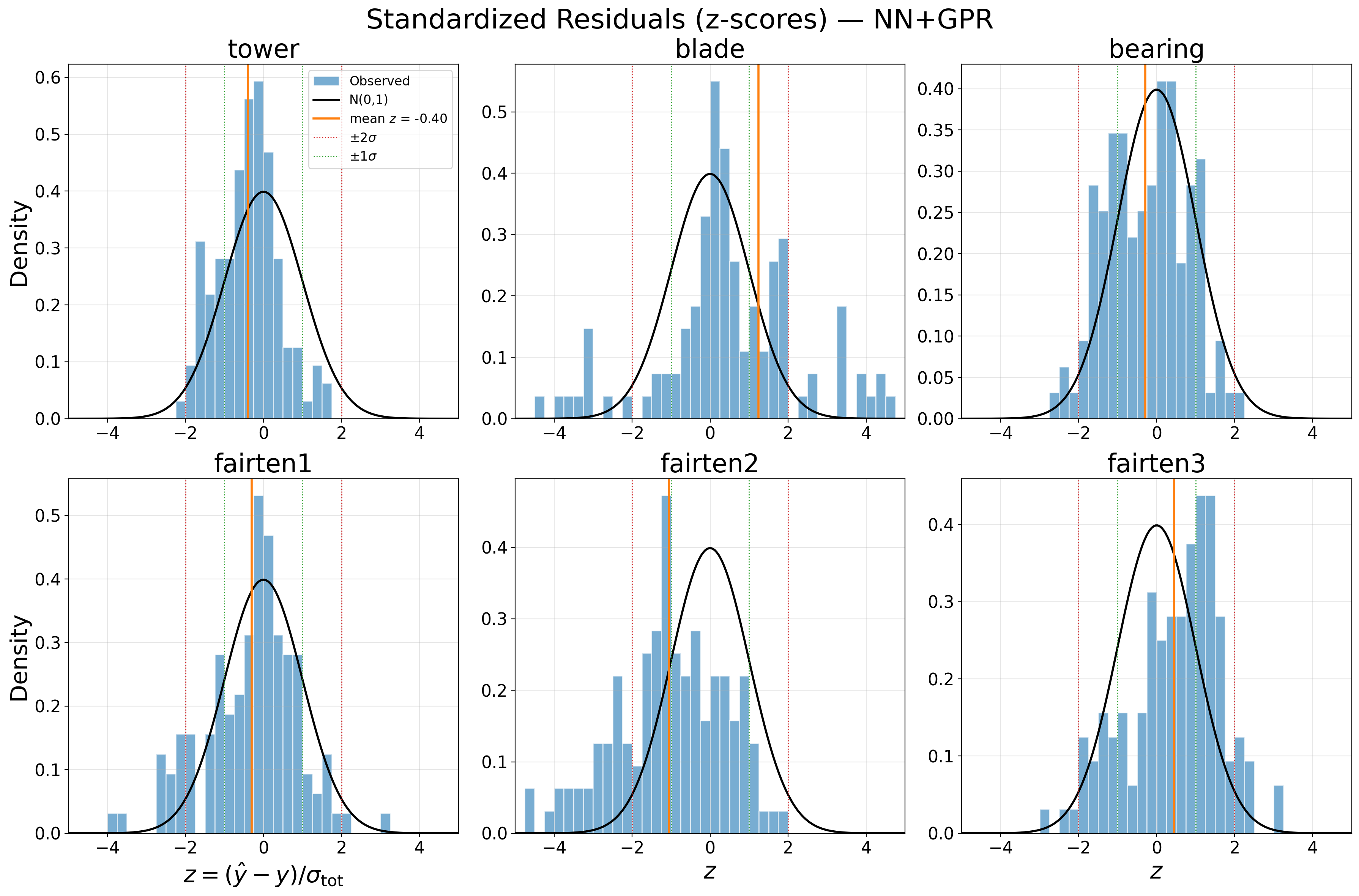}
\captionsetup{skip=2pt}
\caption{Standardised residual distributions for each DEL component. The black curve represents the standard normal distribution $\mathcal{N}(0,1)$. A well-calibrated model produces a histogram that closely matches the reference curve.}
\label{fig:zscore_hist}
\end{figure}

\subsection{Uncertainty assessment}
\label{subsec:uncertainty_assessment}

The calibration of the uncertainty estimates provided by the FFNN+GPR framework is assessed using the four large wind farm simulations described in Sect.~\ref{subsec:Load_surrogate_model_assessment}, which constitute an independent test set of 128~DEL values (4~layouts $\times$ 32~turbines). For each sample, the predicted total uncertainty $\sigma_{\mathrm{tot}}$ is compared against the actual prediction error $|\hat{y} - y|$, where $\hat{y}$ is the FFNN+GPR prediction and $y$ is the FAST.Farm reference value. Under the assumption that the prediction errors follow a Gaussian distribution, the confidence interval at the $k\sigma$ level should contain approximately 68.3\%, 95.4\%, and 99.7\% of the samples for $k = 1, 2, 3$, respectively.

\begin{figure}[ht]
\centering
\includegraphics[width=1\textwidth]{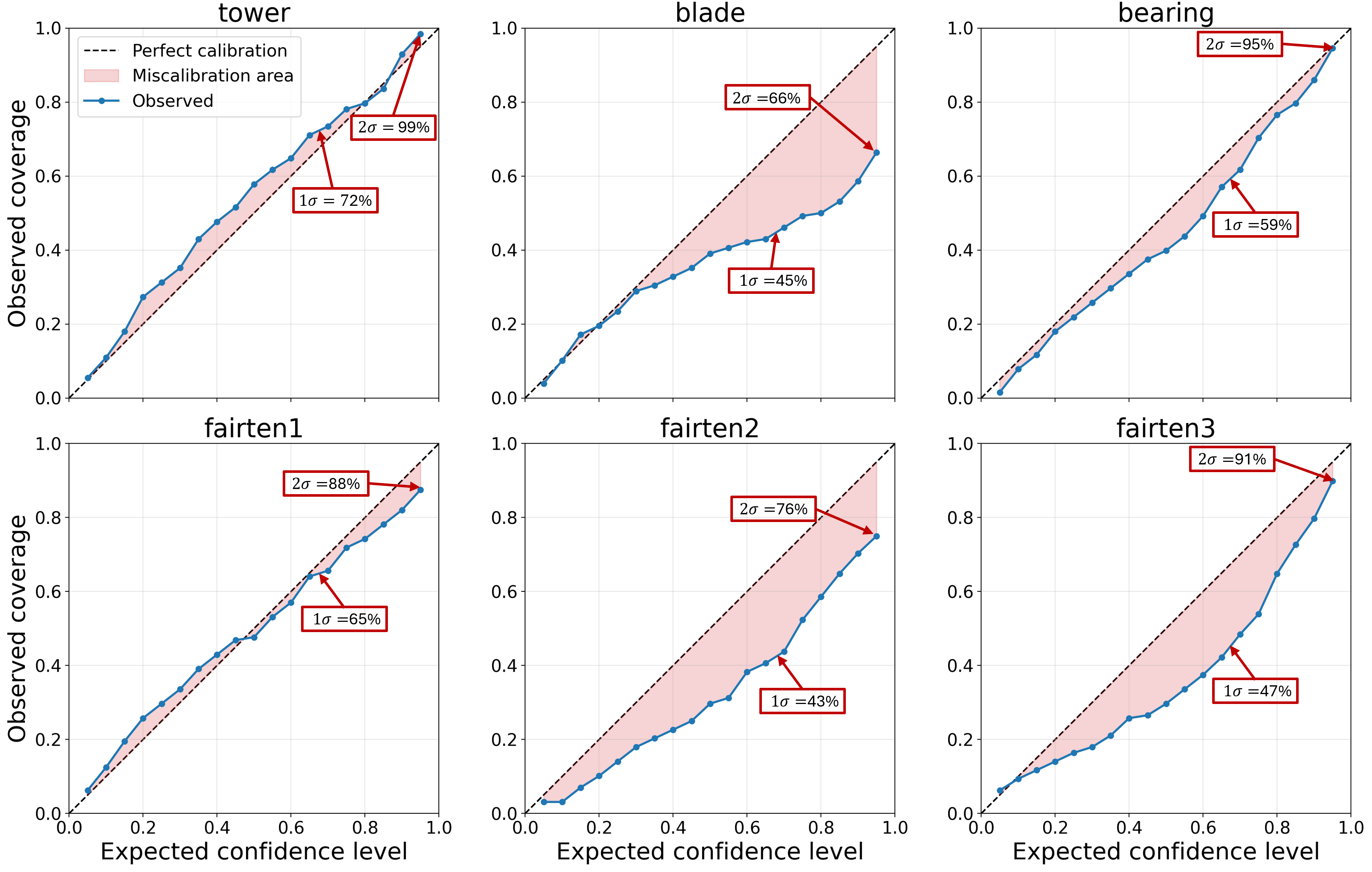}
\captionsetup{skip=2pt}
\caption{Reliability diagrams for each DEL component. The dashed line indicates perfect calibration; the shaded area represents the miscalibration region. Annotated values report the observed coverage at the $1\sigma$ and $2\sigma$ confidence levels.}
\label{fig:reliability_diag}
\end{figure}

\begin{figure}[t]
\centering
\includegraphics[width=1\textwidth]{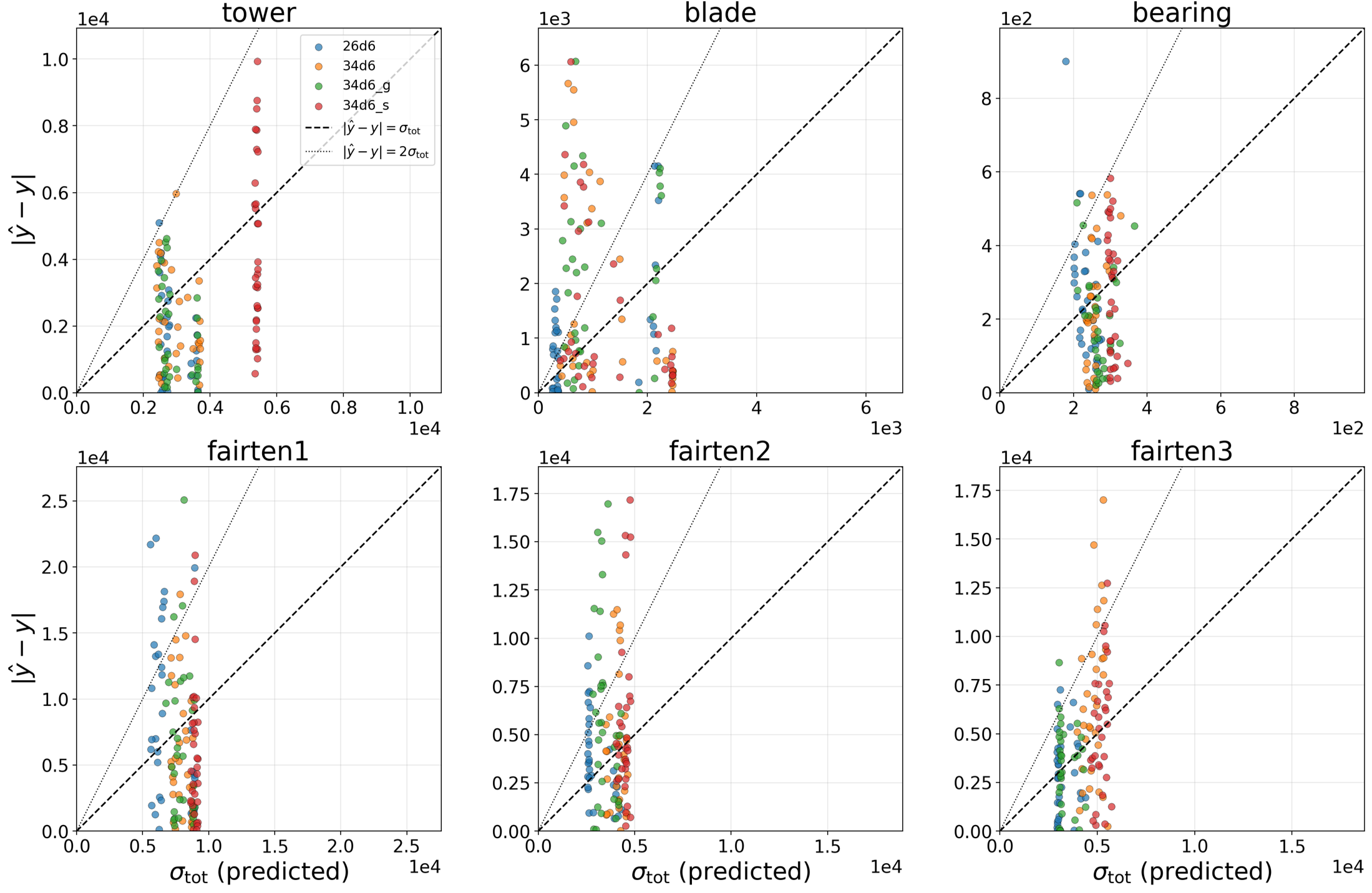}
\captionsetup{skip=2pt}
\caption{Predicted total uncertainty $\sigma_{\mathrm{tot}}$ versus actual absolute prediction error for each DEL component. The dashed and dotted lines correspond to $|\mathrm{error}| = 1\sigma_{\mathrm{tot}}$ and $|\mathrm{error}| = 2\sigma_{\mathrm{tot}}$, respectively. Colours distinguish the four test layouts.}
\label{fig:sigma_vs_error}
\end{figure}

Three complementary diagnostics are employed.
\begin{enumerate}
\item the standardised residuals $z = (\hat{y} - y) / \sigma_{\mathrm{tot}}$ are computed and their distribution is compared in Fig.~\ref{fig:zscore_hist} against the standard normal $\mathcal{N}(0,1)$: a well-calibrated model produces a histogram that closely follows the reference Gaussian curve.
\item reliability diagrams (Fig.~\ref{fig:reliability_diag}) plot the observed coverage against the expected confidence level across the range 5--95\% and perfect calibration corresponds to the diagonal of the plot; the mean calibration error (MCE), defined as the average absolute deviation between observed and expected coverage, quantifies the overall calibration quality.
\item scatter plots of $\sigma_{\mathrm{tot}}$ versus $|\hat{y} - y|$ (Fig.~\ref{fig:sigma_vs_error}) are used to verify whether the predicted uncertainty is informative at the individual-sample level, i.e.\ whether higher predicted uncertainty corresponds to larger actual errors.
\end{enumerate}

Table~\ref{tab:calibration_summary} summarises the coverage results the analysed six DEL components. The tower-base bending moment exhibits excellent calibration, with 71.9\% and 99.2\% coverage at the $1\sigma$ and $2\sigma$ levels, respectively (ideal values: 68.3\% and 95.4\%). The corresponding MCE of 0.045 is the second lowest among all load components, and both the standardised residual histogram and the reliability diagram closely match their theoretical references. This result confirms that, for the tower, the GPR residual correction captures a structured pattern arising from aerodynamic wake interactions, and the associated uncertainty estimate is meaningful.

The blade and yaw bearing DEL components show a different behaviour. For the blade, the FFNN alone already achieves a low MAE of 2.9\%, and the GPR correction degrades the accuracy to 9.8\% while introducing a positive bias of 7.3\%. The calibration is poor (MCE\,=\,0.139), with coverage of only 44.5\% and 66.4\% at the $1\sigma$ and $2\sigma$ levels. The standardised residual distribution is broader and flatter than $\mathcal{N}(0,1)$, indicating that the residual errors exceed the predicted uncertainty bounds. This suggests that, for the blade, the NN-to-FAST.Farm residual lacks a systematic structure that the GPR can learn; the correction is effectively fitting stochastic variability across seeds and wave realisations. The yaw bearing DEL shows a marginal improvement from the GPR (MAE from 9.7\% to 8.9\%) with reasonable calibration at the $2\sigma$ level (95.3\%) but slight undercoverage at $1\sigma$ (59.4\%).

The fairlead tensions exhibit the highest absolute uncertainties and the most challenging calibration behaviour. Fairlead~1 achieves the lowest MCE (0.039) among all components, but its $2\sigma$ coverage (87.5\%) remains below the ideal value, indicating that a subset of samples produces errors that exceed even the $2\sigma$ bounds. These outliers, visible in Fig.~\ref{fig:sigma_vs_error}, are concentrated in the 26d6 layout and are likely related to mooring load sensitivity to wave--current interactions that are not fully captured by the input features. Fairlead~2 represents the worst-calibrated component (MCE\,=\,0.169), with a systematic negative bias of $-16.3$\% and coverage of only 43.0\% and 75.8\% at $1\sigma$ and $2\sigma$. Fairlead~3 shows similar undercoverage at $1\sigma$ (46.9\%) but recovers at $2\sigma$ (90.6\%), with a positive bias of $+10.8$\%.

The $\sigma_{\mathrm{tot}}$ versus $|\mathrm{error}|$ scatter plots (Fig.~\ref{fig:sigma_vs_error}) reveal that, for most components, the predicted uncertainty shows limited correlation with the actual error magnitude: the points tend to cluster in vertical bands rather than spreading along the diagonal. This indicates that the model assigns similar uncertainty levels across different operating conditions, without effectively distinguishing between easy and difficult prediction scenarios. The tower is a partial exception, where two distinct clusters of $\sigma_{\mathrm{tot}}$ values are visible, reflecting the ability of GPR to assign higher uncertainty to samples that lie farther from the training data in the input feature space.

\begin{table}[ht]
\centering
\caption{Uncertainty calibration summary for the FFNN+GPR framework, evaluated on 128 independent test samples. Coverage indicates the percentage of samples for which the absolute prediction error falls within the $k\sigma_{\mathrm{tot}}$ confidence interval.}
\label{tab:calibration_summary}
\begin{tabular}{l c c c c c}
\hline
Component  & Coverage $1\sigma$ [\%] & Coverage $2\sigma$ [\%] & Coverage $3\sigma$ [\%] & MCE [--] & Bias [\%] \\
\hline
Ideal      & 68.3 & 95.4 & 99.7 & 0 & 0 \\
\hline
Tower      & 71.9 & 99.2 & 100.0 & 0.045 & $-4.4$  \\
Blade      & 44.5 & 66.4 &  70.3 & 0.139 & $+7.3$  \\
Bearing    & 59.4 & 95.3 &  99.2 & 0.055 & $-2.4$  \\
Fairlead 1 & 64.8 & 87.5 &  97.7 & 0.039 & $-2.6$  \\
Fairlead 2 & 43.0 & 75.8 &  90.6 & 0.169 & $-16.3$ \\
Fairlead 3 & 46.9 & 90.6 &  98.4 & 0.132 & $+10.8$ \\
\hline
\end{tabular}
\end{table}

Overall, the uncertainty quantification framework proves well calibrated for the tower-base bending moment, which is also the load component where the two-stage correction provides the largest accuracy improvement. For the remaining load components, the predicted uncertainty tends to underestimate the actual errors, primarily due to systematic biases that are not captured by the Gaussian uncertainty model. 

These findings suggest that, in a control application, the FFNN+GPR uncertainty should be used with confidence for the tower DEL, while the estimates for fairlead tensions should be interpreted with caution or complemented by additional safety margins.

\subsection{Demonstration of load surrogate model usage in FLORIDyn}
\label{subsec:demo}

\begin{figure}[ht]
\centering
\includegraphics[width=1\textwidth]{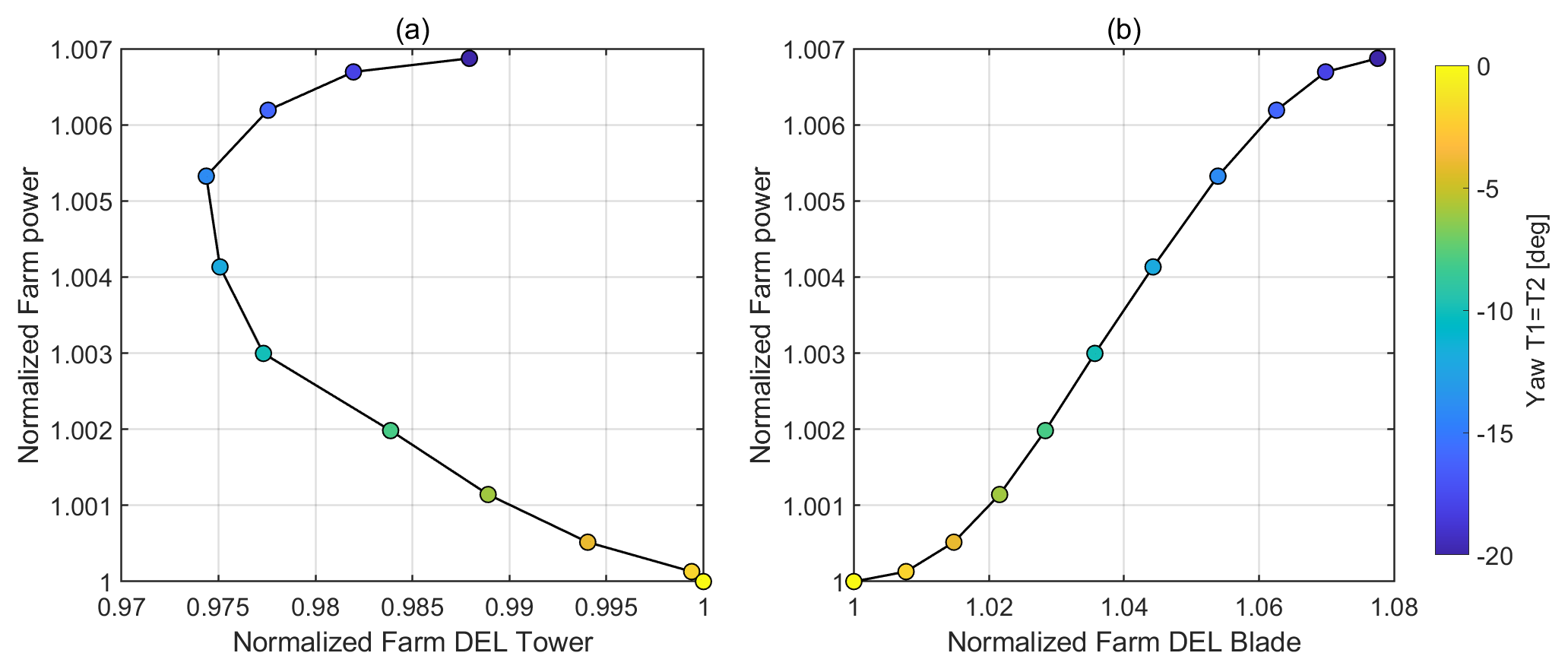}
\captionsetup{skip=2pt}
\caption{Normalized farm-level DEL and power for the three-turbine floating layout: tower (left) and blade (right), for different yaw misalignment magnitudes imposed on the two upstream turbines.}
\label{fig:demo}
\end{figure}

Figure \ref{fig:demo} presents the outcome of the demonstration described in Section \ref{sec:FLORIDyn_demonstration}. The surrogate model employed here follows the architecture described in Section \ref{sec:model_architechture}: the tower DEL is predicted by the FFNN + FFNN pipeline, where a first feed-forward neural network predicts the DEL and a second one estimates the residual correction, while the blade DEL is predicted by a single FFNN without residual correction. The wind farm layout consists of three aligned wind turbines, with a wind speed of 9 $m/s$, a TI of 6\% and a wave height and wave period of 2 meters and 9 seconds, respectively. Each marker in the graphs corresponds to a yaw configuration; the reported metric is the farm-level DEL, obtained by summing the DELs of the three turbines at each simulation step (time step of $4$~s) and averaging the resulting signal over a $3600$~s window. The colorbar encodes the yaw misalignment imposed on the two upstream turbines (T1 = T2), while T3 is kept aligned with the incoming wind.

The blade and tower loads exhibit clearly different trends. For the blade, the load trend is monotonic: as the yaw magnitude increases, the farm power increases due to the wake being steered away from the downstream rotors, but the blade DELs also grow because T2 and T3 operate in higher wind speeds and experience larger cyclic loads. The non-yawed configuration ($\gamma = 0^\circ$) lies at the low-load, low-power corner, while the strongly yawed configuration ($|\gamma| = 20^\circ$) sits at the high-load, high-power corner. In this case a trade-off between power gain and structural loads is direct: any increase in one comes at the cost of the other.

The tower response is more nuanced and exhibits a non-monotonic behavior. For small yaw magnitudes ($|\gamma|$ between $0^\circ$ and $\approx 10^\circ$), the tower DEL initially decreases while the farm power increases, indicating a favorable region where wake steering improves both power and loads simultaneously. Beyond $|\gamma| \approx 12^\circ$, the trend reverses: further increasing the yaw magnitude keeps improving the farm power, but the tower DEL starts to grow. This non-monotonicity highlights the added value of a load-aware controller: a purely power-oriented strategy would push the yaw to its maximum magnitude, whereas a multi-objective formulation that considers loads would identify the knee of the front (around $|\gamma| \approx 10^\circ$ to $12^\circ$) as the most attractive operating point.

\subsection{Surrogate model inference times}
\label{subsec:time_costs}
The computational times presented in this section refer to a workstation equipped with a 13th Gen Intel Core i7-1360P CPU (2.20\,GHz) and 32\,GB of RAM, running on Windows 11. All computations were performed on CPU, without GPU acceleration.
The inference times of the surrogate model are reported for a case of 32 turbines. The first-stage FFNN requires approximately 0.2--0.3\,ms corresponding to less than 0.01\,ms per turbine. The FFNN residual model adds a further 0.13\,ms per batch, resulting in a total inference time lower than 0.5\,ms for the full surrogate model. 

The GPR requires approximately 160\,ms for the 32 turbines, corresponding to roughly 5\,ms per turbine --- a cost that is entirely acceptable for a single inference call. This should, however, be considered carefully when the surrogate model is queried repeatedly within the same control cycle, for instance during a yaw optimization procedure such as Serial Refine, where the cost function may be evaluated tens to hundreds of times. In such cases, the cumulative inference time can become a relevant bottleneck depending on the available control loop time budget: for a budget of, e.g., 15\,s, about 93 evaluations could be performed before exceeding it, excluding the additional time required for the power computation. When this is a practical concern, the FFNN+FFNN framework offers a straightforward alternative, with a total inference time below 0.5\,ms per evaluation, remaining well within the control loop budget regardless of the number of candidates evaluated.

A similar consideration applies to large ensemble evaluations, where uncertainty quantification requires running many realizations of the surrogate --- for instance, to propagate input uncertainty or to compute confidence intervals over the predicted DELs. As an example, an ensemble of 30 realizations run sequentially would require approximately $30 \times 160\,\text{ms} = 4.8\,\text{s}$ for the FFNN+GPR. While this remains compatible with most control update frequencies, the FFNN+FFNN alternative can again be adopted whenever faster ensemble evaluation is required.

\section{Discussions}
\label{sec:discussions}
The surrogate model achieved errors generally below 15\% for the tower base, blade root, yaw bearing, and fairlead~1 across all the investigated operating conditions. This accuracy is comparable to that reported by physics-based digital twin approaches for floating offshore wind turbines, which typically achieve tower DEL errors in the range of 10\%--15\% under numerical validation conditions~\cite{Branlard2024}. However, residual-based correction strategies showed clear limitations when applied to operating conditions outside the training distribution: the GPR residual model exhibited degraded accuracy under out-of-distribution conditions, especially for the blade component, whereas the standalone FFNN showed more robust generalization. Both the FFNN+FFNN and FFNN+GPR frameworks achieved inference times compatible with real-time applications, though with different practical scopes. The FFNN+FFNN approach, requiring less than 0.5\,ms for a full 32-turbine prediction, is suitable for applications demanding multiple surrogate evaluations per control cycle, such as real-time yaw optimisation or ensemble-based uncertainty quantification. The FFNN+GPR framework, with its higher inference time of approximately 160\,ms, remains compatible with real-time applications depending on the controller update frequency and the number of evaluation needed.

Although this paper presents the methodology using only data generated in a simulated environment, generating representative datasets with FAST.Farm is a complex task, and it is difficult to build a training dataset that spans a sufficiently wide range of operational conditions. Indeed, the obtained results and comparison analyses indicate how the large space of possible environmental conditions, coupled with yaw angle control, can be difficult to model accurately, especially in the large wind farm case with a freestream wind speed of 8 m/s and a wind direction of 8$^\circ$. This different operating condition was voluntarily chosen in order to stress the capability of the load surrogate model to generalize over conditions significantly different from those used during training.

Some considerations regarding the evolution of the surrogate model towards a product suitable for industrial applications can be made. The model could be considerably enhanced by incorporating data from real operating environments after the construction of the wind farm, using the model trained on simulated data only as a starting point. The use of real operational data would substantially increase both the amount and diversity of training data. Unlike simulation-based datasets, which are generated from a predefined set of combinations of wind, turbulence, wave, and operational parameters, each 10-minute SCADA record corresponds to an actual operating condition experienced by the wind farm. Over the course of a year, this would result in a large collection of representative operating scenarios and associated DEL values, enabling the surrogate model to be retrained using data that better reflect real-world conditions.

Another possible approach for bringing the surrogate model to operation, considering the quality of the results obtained from the first-stage model, could be the implementation of only the first stage model within a real wind farm, while using GPR or FFNNs to train the correction model for the measured wind farm response, also taking into account the geometry and operational state of the plant. However, this approach would also present some challenges.
The collection of training data from real operation may require the deliberate exploration of control settings that are suboptimal for power production or structural loading. Although these operating points are valuable for improving the model's generalization capability, their use in a commercial wind farm could be difficult to justify due to the associated performance penalties.

Alternatively, a different solution would be to train and validate the surrogate model using wind tunnel experiments. In this environment, the possibility of collecting data is significantly enhanced due to the compressed time scale. An example is the GVPM facility at Politecnico di Milano, where the real-time scale ratio is approximately 1:60. This would provide the possibility of generating high-frequency data in a high-fidelity environment, reducing the equivalent DEL data generation interval from 10 minutes to approximately 10 seconds. A limitation of this approach is that wind tunnel experiments are conducted in a simplified environment compared to a real wind farm.

\section{Conclusions}
\label{sec:conclusions}
In this article, a new load surrogate model for floating offshore wind turbines has been presented. The model is built upon mid-fidelity simulations generated using the OpenFAST–FAST.Farm framework. The proposed methodology consists of two main regression stages: a first stage trained on a large dataset of single wind turbine simulations under high-turbulence inflow conditions, and a second stage incorporating wind farm geometry and operational features. Two formulations of the second stage have been proposed and compared: a Gaussian Process Regression (GPR) and a feedforward neural network (FFNN). The GPR offers greater interpretability through its kernel hyperparameters --- in particular the length scales, which quantify the relative influence of each input feature on the residual correction --- and natively provides uncertainty estimates through its posterior variance. The FFNN, while acting as a black-box model with limited interpretability, achieves comparable prediction accuracy at a substantially lower training and inference cost, making it more suitable for real-time control applications requiring multiple surrogate evaluations per control cycle.

The surrogate model accuracy has been evaluated on the training and validation datasets – consisting of 2700 simulations of 4 wind turbines, for a total amount of 10800 DEL values – and also through four large-wind farm case aimed at assessing the generalization capability of the surrogate model under operating conditions different from those included in the training dataset. 
The obtained results demonstrated that the proposed surrogate model is capable of estimating DELs associated with the main load components of a floating wind turbine with sufficiently low error and low inference time which are suitable for real-time applications such as wind farm closed-loop control or integrity monitoring.

Furthermore, the proposed methodology could be extended to account for additional wind farm control strategies beyond wake steering. In particular, derating and induction control represent promising alternatives that may be adopted to achieve different operational objectives, such as load mitigation or increased robustness under uncertain inflow conditions. The surrogate model could therefore be retrained to include these control actions among its inputs, enabling the evaluation of their impact on both power production and structural loads. Such an extension would broaden the applicability of the framework and support the development of more advanced multi-objective wind farm control strategies. In light of this discussion, future work could focus on improving the dataset by enriching it with additional targeted simulations capable of enhancing the model generalization over a wider range of environmental conditions. Finally, including platform motion characteristics – such as peak amplitude and frequency – among the features of the surrogate model may improve DEL inference accuracy, particularly for fairlead loads. Therefore, developing a strategy to efficiently provide these inputs to the residual model could be crucial to improve the accuracy and robustness of fast DEL computation.

Overall, this work represents a first step towards the development of 
computationally efficient load-aware tools for floating offshore wind 
farms. Although preliminary, the obtained results are promising and 
suggest that the proposed surrogate model could serve as a building block 
for future load-aware optimization and control strategies, contributing 
to the design and operation of the next generation of floating wind farms.

\section*{Acknowledgements}
F.B. acknowledges Eni S.p.A. for funding his PhD scholarship

\section*{Code availability and dataset}
The code is available at "https://github.com/fbelli997/Load-surrogate-model-for-FOWTS-Floating-Offshore-Wind-Turbines-"

\bibliographystyle{unsrt}
\bibliography{DD_model_references}

\end{document}